\documentclass[prb,showpacs,twocolumn,superscriptaddress,longbibliography]{revtex4-1}

\usepackage{amsmath,graphicx,latexsym,bm}

\usepackage[usenames,dvipsnames]{color}

\begin{document}

\title{Unified \emph{ab initio} formulation of flexoelectricity and strain-gradient elasticity}

\author{Massimiliano Stengel}
\affiliation{ICREA - Instituci\'o Catalana de Recerca i Estudis Avan\c{c}ats, 08010 Barcelona, Spain}
\affiliation{Institut de Ci\`encia de Materials de Barcelona 
(ICMAB-CSIC), Campus UAB, 08193 Bellaterra, Spain}

\date{\today}

\begin{abstract} 
The theory of flexoelectricity and that of nonlocal elasticity 
are closely related, and are often considered together when modeling
strain-gradient effects in solids.
Here I show, based on a first-principles lattice-dynamical analysis, that their 
relationship is much more intimate than previously thought, and their consistent
simultaneous treatment is crucial for obtaining correct physical answers.
In particular, I identify a \emph{gauge invariance} in the theory, whereby
the energies associated to strain-gradient elasticity and flexoelectrically
induced electric fields are individually reference-dependent, and only
when summed up they yield a well-defined result.
To illustrate this, I construct a minimal thermodynamic functional 
incorporating strain-gradient effects, and establish a formal link
between the continuum description and \emph{ab initio} phonon dispersion
curves to calculate the relevant tensor quantities. 
As a practical demonstration, I apply such a formalism to bulk SrTiO$_3$, 
where I find an unusually strong contribution of nonlocal elasticity, 
mediated by the interaction
between the ferroelectric soft mode and the transverse acoustic branches.
These results have important implications towards the construction of
well-defined thermodynamic theories where flexoelectricity
and ferroelectricity coexist.
More generally, they open exciting new avenues for the implementation
of hierarchical multiscale concepts in the first-principles simulation
of crystalline insulators.
\end{abstract}

\maketitle

\section{Introduction}

Flexoelectricity, the polarization response of an insulating material
to a strain gradient, has sparked widespread interest in the past few years
as a viable route towards novel electromechanical device concepts.~\cite{pavlo_review,advmat,Yudin-13}
Flexoelectricity is a close relative of piezoelectricity, which 
describes the coupling between strain and polarization. Unlike the
latter, which is present only in crystals that break inversion symmetry,
it is a universal property of all insulators.
The main drawback is that flexoelectricity is negligibly small in
macroscopic samples, and this has limited its practical 
interest until very recently.
The realization that, by downscaling the sample, one can enhance
the effect in a proportion that is roughly inverse with its size,
has motivated the current ``revival''.
A number of interesting functionalities and potential device
applications have been reported recently, 
including the possibility of rotating~\cite{gustau1} or switching~\cite{gustau2} 
the ferroelectric polarization by mechanical means, or of obtaining a 
pseudo-piezoelectric effect that is comparable in magnitude to the existing 
commercial units.~\cite{cross}

Prior to practical exploitation it is crucial, however, to improve
our understanding of how flexoelectricity works at the nanoscale. 
It being a higher-order effect, 
both the theoretical analysis and the interpretation of the experimental results
are highly nontrivial, calling for advanced simulation techniques to cope with 
the many existing subtleties. 
While both first-principles and continuum modeling of flexoelectricity have
undergone impressive progress in the past few years, 
there are strengths and limitations to either approach, suggesting that
only a combined effort will eventually prove itself effective.
Continuum treatments, for example, are best suited at capturing the complexity 
and length scales of a typical flexoelectric measurement, which often
involve nontrivial experimental setups and boundary conditions.
Their main disadvantage is that the quantitative values of the model 
parameters, and sometimes even the specific form of the coupling terms, 
are not always obvious to infer from the existing data, 
physical common sense or basic symmetry considerations.
This is precisely the area where electronic-structure techniques 
could help immensely, by providing a solid microscopic foundation
to the higher-level description; yet, the cross-fertilization between 
the two research areas has remained very limited to date.
Identifying the obstacles that have
prevented such an exchange until now, and devising concrete 
avenues for overcoming them, appears crucial for future progress.

At the most basic level, flexoelectricity can be 
studied via a three-step procedure: first, classical elasticity is
used to solve for the equilibrium strain field in the sample;
next, the polarization due to the strain gradients is computed, and 
finally the Poisson equation of electrostatics is used to
compute the electric potential in some specified electrical 
boundary conditions.
This approach is ideally suited, for example, 
to studying the direct flexoelectric effect, i.e. the 
electrical response to a well-defined mechanical perturbation
of the sample.
Providing quantitative first-principles support to such a
working strategy is now well within reach, as methods~\cite{artlin,artgr} 
for computing the materials-specific values of the bulk flexoelectric
coefficients~\cite{Hong-11,Hong-13,artcalc} and of the relevant 
surface contributions~\cite{artcalc} have
been convincingly demonstrated.

Recent works, however, have emphasized the interest of 
estimating not only the electrical potential, but also 
the \emph{energy} that is associated with flexoelectric
phenomena.
This is necessary, for example, for understanding the impact of 
flexoelectricity on the toughness of materials~\cite{Arias-15} (strain gradients 
are huge in the proximity of a crack tip, suggesting that they may be
crucial for a correct estimation of the energy release rate), or more generally 
for performing a \emph{self-consistent} solution of the electromechanical
problem.~\cite{Arias-14}
%
This goal is much more challenging to achieve, and presents 
several potential difficulties that need to be carefully
considered prior to practical implementation.

The first concern is, of course, ensuring that a
bulk thermodynamic functional is \emph{well defined}, e.g. it
should be immune to the known reference potential 
dependence~\cite{adp,artlin} that characterizes the flexoelectric 
tensor components.
In a nutshell, the loss of periodicity that a strain gradient entails forces 
us to abandon the notion of a ``universal'' macroscopic electric
field, and replace it with the more elusive concept of 
\emph{deformation potential};~\cite{bardeen-50} the latter depends on the (arbitrary) 
choice of the band feature that is taken as a reference, and is therefore nonunique.
Such an ambiguity constitutes
a clear problem at the moment of incorporating flexoelectric
effects in a thermodynamic functional: An obvious consequence, 
for example, is that the Maxwell energy of the electric fields 
generated by a strain gradient is no longer a well-defined 
physical quantity.

A second source of concern is making sure that the thermodynamic
functional contains all the necessary ingredients for a 
realistic description of the physical properties of interest. 
In this context, several independent groups~\cite{Arias-14,Purohit-14} have advocated
the inclusion of \emph{strain-gradient elasticity}~\cite{Mindlin-68,Askes-11} (SGE)
in flexoelectric models. 
SGE has gained increasing popularity in recent years 
as a nonlocal correction to classical elasticity that
is, in principle, able to capture mechanical size effects 
at the nanoscale.
Its dependence on the strain gradient squared is of the same 
order as the Maxwell energy of the flexoelectrically generated
electric fields (the latter are linear in the strain gradient,
and the electrostatic energy depends quadratically on them),
suggesting that these two terms should indeed be treated together.
Unfortunately, the fundamental knowledge of SGE is 
to date very limited. Its practical use in continuum
models involving flexoelectricity has mostly been motivated 
by stability concerns,~\cite{Arias-15} while comparatively
little has been done towards implementing a materials-specific 
treatment of the corresponding physical constants. 

To gain a quantitatively accurate description of SGE,
extracting the relevant coefficients from 
\emph{ab initio} electronic structure simulations appears
to be an excellent idea, particularly in light of the 
experimental difficulties at estimating their values
with an acceptable degree of accuracy.
In this context, the pioneering work of Maranganti and 
Sharma~\cite{Maranganti-07,Maranganti-07b} deserves a special mention. 
These authors developed a 
lattice-dynamical framework to compute the SGE 
tensor components from first-principles, and reported
results for a reasonably wide range of materials including 
metals, semiconductors and insulators.
While their conclusions were 
skeptical regarding the relevance of SGE for 
nanotechnologies in general, there are several good reasons
to revisit the problem in a more fundamental framework.
%
Indeed, there are many convincing indications that 
SGE may be strongly enhanced by flexoelectric couplings:
Axe {\em et al.}~\cite{Axe-70} demonstrated long ago that the 
presence of a ``soft'' optical phonon (as is typical in
ferroelectric materials) may produce an anomalous dispersion 
of the transverse acoustic branch, and similar arguments were
recently invoked to explain the antiferroelectric transition
in PbZrO$_3$~\cite{Tagantsev-12}.
Since SGE is associated precisely with the dispersion of the
acoustic branches, it is reasonable to expect that nonlocal elastic
effects may be particularly strong in such materials.
Unfortunately, the database of crystalline solids that were considered 
in Ref.~\onlinecite{Maranganti-07} did not contain any ferroelectric 
perovskite, thus a quantitative verification of these speculations 
is still missing.
Even at the qualitative level, there is a clear need to establish
a sound theoretical formalism describing both flexoelectricity and
SGE from a fundamental perspective, and clearly relating either
macroscopic property to the microscopic physics of the insulating
crystal.

Here I propose a general strategy to address the aforementioned 
questions by constructing a continuum theory,
incorporating flexoelectricity and other strain-gradient effects, 
\emph{directly from first principles}, via a number of
well-defined, controlled approximations.
%
A long-wave expansion of the dynamical
matrix of the crystal around the Brillouin zone center, where the continuum
fields are associated with the transverse lattice modes therein,
naturally provides such a framework.
By appropriately choosing the order (in powers of the wavevector $q$)
at which the Taylor expansion is truncated, one can readily decide,
in an unbiased manner, what physical properties to include or
exclude from the model, and yet rest assured
that the higher-level description is still \emph{exact} (i.e.
of full \emph{ab initio} accuracy) and
\emph{well-defined} at the targeted length scales.

To demonstrate these ideas in practice, I will show that bulk SrTiO$_3$ 
is an excellent model system, and will use it to discuss to a number of key topics, 
including: the relationship between flexoelectric and nonlocal elastic effects;
the role of the long-range electrostatic interactions, especially
in light of the aforementioned reference-potential dependence; 
some peculiarities of (incipient) ferroelectric materials,
where strain-gradient effects are expected to be particularly strong.
I find that: (i) The energies associated to strain-gradient elasticity 
and flexoelectricity are both reference-dependent in the sense specified
in Ref.~\onlinecite{adp}, but their respective arbitrariness cancels out when the
two terms are summed up -- explicit inclusion of \emph{both} is therefore
crucial for ensuring that the functional is well defined;
(ii) The flexoelectric contribution to the SGE energy is systematically
\emph{negative}, i.e. it results in a softening of the elastic
response at short length scales; (iii) The SGE energy \emph{diverges}
in a vicinity of a ferroelectric transition, where
the coupling between the transverse acoustic and optical
soft-mode branch may lead to a markedly nonlocal elastic response.

To substantiate the above statements, I introduce the 
concept of \emph{energy flexocoupling tensor}, which describes
the coupling between a macroscopic strain gradient and 
an arbitrary zone-center optical mode, and report 
a complete calculation of its independent entries in 
bulk SrTiO$_3$.
This, together with the ``frozen-ion''~\cite{Hong-11} 
flexoelectric and strain-gradient elasticity tensors, 
provides complete information to describe both
flexoelectric and SGE effects in bulk SrTiO$_3$, both
at the electronic and lattice-mediated levels.
In addition, I use the formalism developed here to address
a number of related subtleties, regarding for example
the static or dynamic nature of the SGE and flexocoupling
constants, and whether both coexist as separately measurable 
contributions.~\cite{Kvasov-15}
I will show that, in this respect, the strain-gradient elasticity tensor 
behaves similarly to the flexoelectric~\cite{artlin} tensor: it is an 
intrinsically dynamic object, and hence its individual components generally 
\emph{depend} on how the mass density of the crystal is distributed among the 
basis atoms of the primitive cell. 
Yet, for any deformation field at rest, such mass dependence  
cancels out due to the mechanical equilibrium condition, yielding
``effective'' SGE coefficients that are \emph{static} quantities,
as one would expect.~\cite{artlin}
Finally, I shall briefly discuss the thermodynamic stability of 
continuum models involving strain-gradient effects, demonstrating how the
formalism developed here naturally provides alternative routes to 
addressing some long-standing~\cite{Askes-11} issues in this context.

This work is organized as follows: In Sec.~\ref{sec:general} I shall 
introduce some general concepts regarding the continuum energy functional
and its mapping onto the discrete lattice model. In Sec.~\ref{sec:theory}
I shall explicitly derive the coupling terms via a long-wave perturbative
expansion of the harmonic force constants. In Sec.~\ref{sec:results} I
shall present the numerical results for SrTiO$_3$. In 
Sec.~\ref{sec:discuss} and Sec.~\ref{sec:conclusions},
I shall discuss the aforementioned stability issues and draw some 
general conclusions.

\section{General background}

\label{sec:general}

\subsection{Continuum thermodynamic functional}

\label{sec:continuum}

Classical elasticity is commonly described in terms of the following Lagrangian
density,
\begin{equation}
\mathcal{L}({\bf u},\dot{\bf u}) = 
   \frac{\rho_{\rm M}}{2} |\dot{\bf u}|^2 - \frac{1}{2} \bm{\varepsilon} \cdot \bm{\mathcal{C}} \cdot \bm{\varepsilon},
\label{lelas}
\end{equation}
where $\bm{\varepsilon}$ is the symmetrized strain tensor, $\bm{\mathcal{C}}$ 
is the fourth-rank elastic tensor, ${\bf u}({\bf r})$ is the displacement field and
$\rho_{\rm M}$ is the mass density of the crystal.
In order to describe nonlocal effects, which may become important at very short 
length scales, strain-gradient corrections have been proposed, typically in the following form,
\begin{equation}
E^{\rm SGE} = \frac{1}{2} \nabla \bm{\varepsilon} \cdot {\bf H} \cdot \nabla \bm{\varepsilon},
\label{eqsge}
\end{equation}
where ${\bf H}$ is the sixth-order strain-gradient elasticity (SGE) tensor,
also known as ``hyperelastic'' tensor.
This formulation is good enough for a metal, but necessarily incomplete for an 
arbitrary insulator: Flexoelectricity states that strain gradients are 
universally associated with an electric polarization,
\begin{equation}
P_\alpha = \mu^{\rm II}_{\alpha \lambda, \beta \gamma} \frac{\partial \varepsilon_{\beta \gamma}}{\partial r_\lambda},
\label{eqp1}
\end{equation}
where $\bm{\mu}^{\rm II}$ is the total type-II~\cite{artlin} flexoelectric tensor (including
electronic and lattice-mediated contributions). 
This means that, when dealing with 
strain gradient elasticity, additional electrostatic terms are necessary to account
for the Maxwell energy of the macroscopic longitudinal fields
\begin{equation}
E^{\rm M} = \frac{1}{2} \frac{|{\bf P_\parallel}|^2}{\epsilon_0 \epsilon},
\label{Maxwell}
\end{equation}
where ${\bf P_\parallel}$ stands for the irrotational component of ${\bf P}$, 
$\epsilon$ is the static dielectric constant (assuming it to be isotropic for 
simplicity), and $\epsilon_0$ is the permittivity of vacuum.
As ${\bf P}$ is linear in the strain gradient amplitude, $E^{\rm M}$ goes
like the strain gradient \emph{squared}, i.e. it is of the same order as
$E^{\rm SGE}$.
In a way, the relationship between strain-gradient elasticity and flexoelectricity
parallels that existing between classical elasticity and piezoelectricity.
In both cases, the stiffness to a mechanical deformation is influenced by the
electrical boundary conditions, and such dependence boils down to the Maxwell
energy associated to the open-circuit electric fields.
The ``flexoelectric energy'', from this perspective, is just one of the 
contributions to the SGE energy, pretty much the same way as the 
``piezoelectric energy'' (i.e. the
direct copuling of the zone-center optical modes to the strain) contributes
to the elastic tensor.
Thus, just like in the case of the elastic tensor in a piezoelectric material,~\cite{Wu-05}
one can define different versions of the SGE tensor depending on the electrical
boundary conditions that are applied to the crystal. In the remainder of this work,
unless otherwise specified, we shall assume that ${\bf H}$ is defined under
\emph{short-circuit} boundary conditions; this is important for reasons that shall become
clear shortly.

The above functional contains the minimal amount of physical ingredients to describe, 
at the same time, flexoelectricity and strain-gradient elasticity, provided that the 
deformations are smooth enough (i.e. that higher-order gradients of $\bm{\varepsilon}$ can 
be neglected) and their amplitude is small (linear limit).
Of course, more sophisticated choices are possible, e.g. by explicitly treating
additional fields (together with the mechanical deformation) as independent 
dynamical variables in the Lagrangian density. 
The most obvious strategy in this context would be to explicitly treat the ferroelectric
``soft-mode'', which might be unavoidable in most systems of practical interest. 
(Ferroelectric perovskites are, 
among crystalline materials, the most promising and well
studied from the point of view of flexoelectricity.)
As we shall see, the specific choice of the target functional is 
largely irrelevant to our scopes: This work will mostly focus on 
how to extract the basic ingredients (in the form of coupling
coefficients) from an \emph{ab initio} model -- these can
easily be incorporated later in a variety of continuum 
Lagrangians.
To avoid unnecessary complications, I shall stick to the formulation 
described above throughout this work, and briefly discuss some useful 
alternatives in Sec.~\ref{sec:discuss}.

%
%
%
%
%

\subsection{Reciprocal-space formulation}

\label{sec:recip}

In order to bring the continuum functional into a 
form that is directly compatible with \emph{ab initio} 
lattice dynamics, it is convenient to Fourier-transform the displacement field ${\bf u}$ as follows,
\begin{equation}
{\bf u} ( {\bf r}) = \frac{1}{\sqrt{(2\pi)^3}} \int d^3 q \, {\bf U}({\bf q}) e^{i {\bf q \cdot r}}.
\end{equation}
The Lagrangian density can be then written in reciprocal space as 
(for clarity, I use Latin indices for the wavevector components and 
Greek indices otherwise)
\begin{eqnarray}
\tilde{\mathcal{L}} ({\bf U}, \dot{\bf U}) &=& \, \, \, \frac{\rho_{\rm M}}{2} |\dot{\bf U}|^2 - 
\frac{1}{2} U_\alpha U_\beta q_i q_j c_{\alpha \beta,  i j}   \nonumber \\
&& - \frac{1}{2} U_\alpha U_\beta q_i q_j q_k   q_l h_{\alpha \beta, ijkl} \nonumber \\
&& - \frac{1}{2} \frac{|{\bf q \cdot P}|^2}{\epsilon_0 \epsilon q^2},
\label{eq:lq}
\end{eqnarray}
where
\begin{equation}
{P}_\alpha  = - \mu^{\rm I}_{\alpha \beta, j k} U_\beta q_j q_k.
\end{equation}
and the reciprocal-space coupling tensors are related to the real-space
ones via a symmetrization of the indices,
\begin{eqnarray}
c_{\alpha \beta,  i j}           & = & {\rm sym}_{(ij)} \, \mathcal{C}_{\alpha i, \beta j}  , \\
\mu^{\rm I}_{\alpha \beta, ij}  & = & {\rm sym}_{(ij)} \, \mu^{\rm II}_{\alpha i, \beta j} , \\
h_{\alpha \beta, ijkl}           & = & {\rm sym}_{(ijkl)} \, H_{\alpha ij,\beta kl}.
\end{eqnarray}
The first thing that one can note from the above formulas is
that classical elasticity is an $\mathcal{O}(q^2)$ effect, while
both the electrostatic and SGE energy terms are $\mathcal{O}(q^4)$.
This is consistent with the observation that I have made in the previous Section,
that the Maxwell energy of the flexoelectric fields and the energy associated with
SGE effects should be regarded as intimately related and of comparable
importance.
%
From the technical point of view, this implies that some specific precautions
need to be taken when calculating ${\bf h}$ from first principles.
Given the nonanalytic character of electrostatic interactions (due to the $q^2$ 
factor at the denominator)
it is of primary importance to define (and calculate) ${\bf h}$ in \emph{short-circuit} 
electrical boundary conditions, otherwise a tensorial expression such as that of Eq.~(\ref{eqsge})
would not be possible.~\footnote{Issues of this kind are, again, well known in the piezoelectric case, 
  where
  the elastic coefficients need to be defined under short-circuit electrical boundary 
  conditions for $\bm{\mathcal{C}}$ to behave as a tensor.~\cite{Wu-05}}
At first sight, this observation appears to be problematic to implement here, 
as the notion of macroscopic electric field is ambiguous in 
presence of strain gradients.~\cite{adp} 
As we shall see in the following, however, such arbitrariness in the definition 
of ${\bf h}$ is \emph{necessary} in order to guarantee that the functional 
as a whole be well defined, as it exactly cancels with the equal (and opposite)
reference dependence that is implicit in the Maxwell term.

\subsubsection{Gauge invariance}

To understand the origin of the reference dependence, 
note that one can always rewrite the flexoelectric tensor
by separating an \emph{isotropic} contribution from the remainder, $\bm{\mu}'$
(I shall assume in the next few equations that $\bm{\mu}$ is 
represented in type-I form and omit the corresponding superscript),
\begin{equation}
\mu_{\alpha \beta, \gamma \lambda} = \frac{V_0 \epsilon_0 \epsilon}{2} 
  ( \delta_{\alpha \lambda} \delta_{\beta \gamma} + \delta_{\alpha \gamma} \delta_{\beta \lambda} ) + 
  {\mu}'_{\alpha \beta,  \gamma\lambda}.
\label{v_0}
\end{equation}
$V_0$ has the dimension of a potential, and is used here to 
emphasize the physical meaning of the new term in Eq.~(\ref{v_0}): 
this is essentially a \emph{relative deformation potential} that 
modifies the definition of the macroscopic electric field.
The longitudinal polarization then reads as
\begin{equation}
{\bf q \cdot P} = -V_0 \epsilon_0 \epsilon q^2 {\bf q \cdot U}  - \mu'_{i \beta, j k} U_\beta q_i q_j q_k.
\end{equation}
As a result, the original Maxwell energy can be rewritten as,
\begin{equation}
\frac{1}{2} \frac{|{\bf q \cdot P}|^2}{\epsilon_0 \epsilon q^2} = 
\frac{1}{2} \frac{|{\bf q} \cdot {\bf  P}'|^2}{\epsilon_0 \epsilon q^2} + \Delta E,
\end{equation}
where the polarization has been redefined as
\begin{equation}
P'_i  = - \mu'_{i \beta, j k} U_\beta q_j q_k,
\end{equation}
and the remaining term is
\begin{equation}
\Delta E = \frac{ \epsilon_0 \epsilon}{2} V_0^2  q^2 ({\bf q \cdot U})^2 - V_0 ({\bf q \cdot U}) \mu'_{i \beta, j k} U_\beta q_i q_j q_k.
\end{equation}
A key point here is that $\Delta E$ is an \emph{analytic} function
of ${\bf q}$, and therefore can be readily reabsorbed into the SGE
energy via a redefinition of the ${\bf h}$ tensor.
This leads to one of the main results of this work: There is
a sort of \emph{gauge invariance} in the combined theory of
flexoelectricity and strain-gradient elasticity in insulators, 
whereby the SGE and electrostatic energies are separately
ill-defined, but their sum is invariant with respect to a 
simultaneous \emph{gauge transformation} of both the $\bm{\mu}$-
and ${\bf h}$-tensors.
In other words, the arbitrariness of the reference, which can be conveniently
rationalized within the theory of deformation potentials,~\cite{adp,vandewalle-89,Resta-DP} 
only affects the way the total energy is partitioned 
between the electrostatic and SGE parts, without affecting the physical answers that one 
extracts from the functional as a whole.
Note that the expression ``gauge invariance'' is loosely borrowed from
electromagnetism, where there also exists a freedom in the choice of the potentials
(scalar and vector) that enter the governing equations, and yet the physically mesurable
quantities are unsensitive to such a choice. The analogy, for the purposes of
the present work, stops here: for example, it is not obvious how to identify a
counterpart of the magnetic field in the context of the electromechanical
effects under study.

An interesting consequence of the above considerations is that in an
\emph{isotropic} medium the electrostatic energy becomes an analytic function
of ${\bf q}$, and therefore can be reabsorbed into the strain-gradient 
squared term.
This implies that the long-ranged part of the flexoelectrically generated 
electric fields is, in fact, entirely related to the \emph{anisotropy} of the 
electromechanical response.

\subsubsection{Symmetrization of the indices}

The symmetrization of the tensor indices that we have performed
when moving from real space to reciprocal space 
has no consequences regarding the flexoelectric and elastic tensors:
In both cases, symmetrization preserves the number of independent 
entries, and the relationship between the symmetrized and unsymmetrized
representations is readily invertible.
(In the flexoelectric case, the two forms of the tensor have been
indicated as ``type-I'' and ``type-II'' in earlier works;~\cite{artlin}
I shall follow the same convention here.)
%
Things differ in the SGE case: In the lowest-symmetry material the ${\bf h}$-tensor
has $6 \times 15 = 90$ independent entries, after taking into account the invariance of 
$h_{\alpha \beta,ijkl}$ under either $\alpha \beta$ exchange or $ijkl$ permutation.
%
This is much smaller than the total number of entries of the real-space
${\bf H}$-tensor, which is $171$ (the strain-gradient tensor has 18 components, 
and ${\bf H}$ can be regarded as a symmetric square matrix).
Thus, contrary to the cases of standard elasticity and flexoelectricity, 
one cannot invert the relationship between reciprocal- and real-space
SGE coefficients.

This fact has sometimes been regarded as a limitation of
the lattice-dynamical method at computing the SGE coefficients.
(A reciprocal-space representation of the Lagrangian density is typically
performed in the context of lattice-dynamical studies.)
To emphasize this apparent difficulty, it has become 
common practice to indicate ${\bf H}$ as the \emph{static} SGE tensor, and
${\bf h}$ as the \emph{dynamic} one.
Such an appellation is, however, prone to confusion:~\footnote{The 
  terms ``static'' and ``dynamic'' may refer to the
  physical nature of a given effect, or to the procedure 
  that one uses to measure or calculate it. In this work I 
  shall use the former meaning. Static properties can
  be studied by dynamical means and viceversa, so the two
  categories do not always overlap.} 
A strain gradient is an inherently dynamic object (for example,
a purely longitudinal gradient of the type $\varepsilon_{11,1}$
cannot be sustained by any conceivable combination of static 
surface loads~\cite{artlin}), so even the purportedly static
${\bf H}$-tensor components have, in fact, a \emph{dynamic}
nature.
(I shall come back to this important point in Sec.~\ref{sec:static}.)
To avoid misunderstandings, in the remainder of this work 
I shall refer to ${\bf H}$ as the ``type-II'' SGE tensor
(it is associated to strain gradients in type-II form),
and to ${\bf h}$ as the ``symmetrized'' SGE tensor.

In order to better understand the relationship between ${\bf H}$
and ${\bf h}$, and the physical nature of the information 
that has been lost upon symmetrization of the indices, 
it is useful to go back to real space, and write the
SGE energy in type-I form (i.e., replace $\nabla \bm{\varepsilon}$
with the second gradient of the displacement field ${\bf u}$).
Via two subsequent integrations by parts, one can
rewrite the energy as a function of ${\bf u}$
and its fourth gradient, plus a number of surface 
terms.
One can then show that the volume contribution only 
depends on the ${\bf h}$-tensor components; in other 
words, by replacing ${\bf H}$ with ${\bf h}$ 
one leaves the governing bulk equations unaltered,
only the boundary conditions change.
Thus, the distinction between ${\bf H}$ and ${\bf h}$ is
rooted, rather than in their static or dynamic nature, in the 
fact that the latter is a purely \emph{bulk property}, while
the former contains additional surface-specific information.

Surface contributions are, of course, important for the description of 
flexoelectric effects in a \emph{finite} object,
even in the thermodynamic limit of a macroscopically thick sample.
We expect that the local piezoelectric and elastic properties of
the boundary, which might markedly differ from those of the 
homogeneous bulk material, will affect the SGE response of a 
finite sample in a qualitatively similar way. 
However, because of their surface-specific nature, one cannot
generally estimate the corresponding physical constants in the 
context of bulk calculations.
In this work we shall restrict our analysis to the bulk part of 
the energy functional, and on physical phenomena (acoustic phonons) 
where surface contributions play no role.
One must keep in mind, however, that to attack a more general class
of deformations, as for example the response of a slab to bending,
careful considerations of the aforementioned surface terms is unavoidable; 
we shall defer their treatment to a future publication.

\subsection{From discrete to continuum}

\label{sec:discrete}

I shall illustrate in the following how the continuum theory 
that has been outlined in the previous Section can be derived via a 
well-defined approximation of the discrete lattice model.
The Lagrangian of a crystalline system can be written as
\begin{equation}
L(u,\dot u) = T(\dot u) - V(u),
\end{equation}
where $u$ represents the displacements of the atoms from their equilibrium locations.
Within the harmonic approximation, the kinetic and potential terms respectively read as
\begin{eqnarray}
T(\dot u) &=& \frac{1}{2} \sum_{l\kappa \alpha} m_\kappa (\dot{u}^l_{\kappa \alpha})^2, \\
V(u) &=& \frac{1}{2} \sum_{l\kappa l' \kappa'} {\bf u}^l_{\kappa} \cdot \bm{\Phi}^{ll'}_{\kappa \kappa'} \cdot {\bf u}^{l'}_{\kappa'}.
\end{eqnarray}
I shall use the convention from now on that $l$ and $l'$ are cell indices,
$\kappa$ and $\kappa'$ are sublattice indices, and $\alpha$, $\beta$, etc. are
Cartesian directions.
$m_\kappa$ is the atomic mass of specie $\kappa$, and $\bm{\Phi}^{ll'}_{\kappa \kappa'}$
is the real-space force-constant matrix of the periodic crystal.

As above, we move to reciprocal space via the following definition,
\begin{equation}
{\bf u}^l_{\kappa} = \frac{\Omega}{(2\pi)^3} \int_{\rm BZ} d^3 q \, {\bf u}_\kappa^{\bf q} \, e^{ i {\bf q} \cdot {\bf R}_l},
\end{equation}
where ${\bf R}_l$ is a Bravais lattice vector indicating the location of the 
$l$-th cell.
One obtains the Lagrangian density in reciprocal space,
\begin{eqnarray}
L &=& \frac{\Omega}{(2\pi)^3} \int_{\rm BZ} d^3 q \, L^{\bf q}, \\
L^{\bf q}(u, \dot u) &=& T^{\bf q}(\dot u) -  V^{\bf q}(u), \\
T^{\bf q}(\dot u) &=& \frac{1}{2} \sum_{\kappa} m_\kappa 
  \dot{\bf u}^{\bf q}_{\kappa} \cdot \dot{\bf u}^{\bf q}_{\kappa}, \\
V^{\bf q}(u) &=& \frac{1}{2} \sum_{\kappa \kappa'} 
  {\bf u}^{\bf q}_{\kappa} \cdot \bm{\Phi}^{\bf q}_{\kappa \kappa'} \cdot {\bf u}^{\bf q}_{\kappa'}.
\end{eqnarray}

We can now move to a normal mode representation, where the 
(mutually coupled) atomic displacements are replaced by a set
of independent harmonic oscillators, whose amplitudes 
$v_j$ are the new independent variables of the problem,
\begin{equation}
L^{\bf q} (v,\dot v) = \frac{M}{2} (\dot v_{j{\bf q}}^2 - v_{j{\bf q}}^2 \omega_{j{\bf q}}^2).
\label{lqv}
\end{equation}
Here $\omega_{j{\bf q}}^2$ are the eigenvalues of the dynamical matrix,
which can be conveniently represented in an operator form,
\begin{eqnarray}
\hat{D}({\bf q}) |j {\bf q} \rangle &=& \omega_{j{\bf q}}^2 | j {\bf q} \rangle, \\
\langle \alpha \kappa | \hat{D}({\bf q}) | \beta \kappa' \rangle &=& 
  \frac{1}{ \sqrt{m_\kappa m_{\kappa'} } } \Phi^{\bf q}_{\alpha \kappa, \beta \kappa'}.
\label{dynm}
\end{eqnarray}
($| j {\bf q} \rangle$ is the $j$-th mode eigenvector at ${\bf q}$; 
$| \alpha \kappa \rangle$ indicates a hypothetical mode where the atom 
$\kappa$ displaces along $r_\alpha$ while the other sublattices
remain still; all bras and kets are assumed to be normalized to unity.)
Note that the mass factor $M$ is, in principle, arbitrary, but 
is most appropriately set as the total mass of the unit cell:
As we shall see shortly, such a choice leads to a direct identification 
of the mode amplitudes $v_{j{\bf q}}$ with the continuum deformation field.
The normal mode amplitudes are related to the atomic displacements via
\begin{equation}
u^{\bf q}_{\kappa \alpha} = v_{j{\bf q}} \sqrt{\frac{M}{m_\kappa}} \langle \kappa \alpha | j {\bf q} \rangle.
\label{ufromv}
\end{equation}
As such a linear relationship between the $u^{\bf q}_{\kappa \alpha}$
and the $v_{j{\bf q}}$ variables exists, what we have done so far is
simply a change of variables, but we really haven't made any explicit
assumption about the static or dynamic nature of the theory.

At this point we are ready to operate an \emph{adiabatic} approximation,
by supposing that, at the energy and time scale of the phenomena under
study, the optical modes are infinitely fast, and can be considered as 
separated from the acoustic branches. 
In other words, the optical modes are always in their equilibrium
state in the instantaneous deformation field provided by the
``heavy'' acoustic modes.
This implies that the required information on the continuum-theory tensors 
has to be sought in the long-wave behavior of the lowest three eigenvalues 
of the dynamical matrix, i.e. those describing the acoustic phonon branches.
In particular, the corresponding tensor components are trivially
related to the long-wave expansion terms of the squared eigenfrequencies 
by a factor of $\rho_{\rm mass} = M/\Omega$, i.e. the mass density of
the crystal. (This is the total mass of the unit cell, $M=\sum_\kappa m_\kappa$, 
divided by its volume, $\Omega$.)
How to expand the dynamical matrix eigenvalues will be explained in the 
next Section.

\section{Lattice-dynamical theory}

\label{sec:theory}

\subsection{Variational formulation}

Considering an acoustic phonon mode with wavevector 
${\bf q} = q \hat{\bf q}$, where the direction $\hat{\bf q}$ 
shall be kept fixed for the time being, in a vicinity of the 
$\Gamma$-point (center) of the Brillouin zone.
Its squared frequency can be written as a constrained variational functional
of the eigendisplacements vector, $| v(q) \rangle$,
\begin{equation}
G(q) = \langle v(q) | \hat{D}(q) | v(q) \rangle - X(q) (\langle v(q) | v(q) \rangle - 1),
\label{constrained}
\end{equation}
where the dynamical matrix operator, $\hat{D}(q)$, is related to the force-constant 
matrix, $\bm{\Phi}^{\bf q}$, as specified in Eq.~(\ref{dynm}).
%
(Note that the nonanalytic terms related to long-range interactions are
included in $\hat{D}(q)$, i.e., this is the full dynamical matrix.)
%
$X(q)$ is a Lagrange multiplier taking care of the normalization constraint --
at the variational minimum it corresponds to the lowest eigenvalue of $\hat{D}(q)$,
\begin{equation}
X(q) = E(q), \qquad \hat{D}(q) | v(q) \rangle = E(q) | v(q) \rangle.
\label{eigen}
\end{equation}
This, in turn, relates to the phonon frequency as $E(q)=\omega^2(q)$.

Before going through the analytical derivations, it is useful to introduce
here the concept of ``mixed electrical boundary conditions'' (MEBC), which was
originally proposed, in the context of flexoelectricity, by Hong and Vanderbilt.~\cite{Hong-11,Hong-13}
It consists in imposing open-circuit conditions along a given spatial direction (which translates
in constraining the corresponding component of the electric displacement field, ${\bf D}$, to zero),
and short-circuit (that is, a vanishing projection of the electric field vector, ${\bf E}$) in the
normal plane.
This regime is crucially important to understand in the context of a long-wavelength
phonon, where MEBC naturally arise along the propagation direction, $\hat{\bf q}$.
(Other physical contexts where MEBCs occur are, e.g., an unsupported slab in vacuum, 
or a parallel-plate capacitor in open circuit.~\cite{fixedd})
In fact, MEBCs are responsible for the strongly nonanalytic behavior of the phonon response 
functions in a vicinity of ${\bf q=0}$; conversely, if we fix the direction $\hat{q}$,
the electrical boundary conditions remain fixed as well, which implies that the 
response becomes a smooth function of the one-dimensional parameter $q$.

\subsection{$2n+1$ theorem and long-wave expansion.}

The dynamical matrix and its eigenvectors can be then expanded as a perturbation 
series in the small parameter $q$,
\begin{eqnarray}
\hat{D}(q) &=& \hat{D}^{(0)} + q \hat{D}^{(1)} + q^2 \hat{D}^{(2)} + \ldots, \\
|v(q)\rangle  &=& |v^{(0)}\rangle + q | v^{(1)} \rangle + q^2 | v^{(2)}\rangle + \ldots.
\end{eqnarray}
By plugging these expansions into the eigenvalue problem of Eq.~(\ref{eigen}) 
one can readily compute $|v^{(n)}\rangle$ for an arbitrary $n$. Such a procedure 
has been pushed in earlier works~\cite{Born/Huang,Tagantsev,Hong-13,artlin} up
to $\mathcal{O}(q^2)$, which is enough to describe both piezoelectricity 
($n=1$) and flexoelectricity ($n=2$).  

Here we are interested, rather than in the eigenvectors, in the 
$q$-expansion of the dynamical matrix eigenvalues.
This can be conveniently obtained by expanding the constrained functional 
$G(q)$, rather than directly $E(q)$.
The advantage is that, by means of the $2n+1$ theorem~\cite{Gonze-95}, one can
systematically construct even-order $G^{(2n)}$ functionals (odd-order terms are 
forbidden by time-reversal symmetry, which is assumed to hold throughout this 
work) that are \emph{variational} in the eigendisplacements, $|v^{(n)}\rangle$.
%
%
%
%
%
%
%
%
%
As the strain and strain gradient
effects show up, respectively, at the first and second order in $q$, one
needs to push the expansion of the energy to second and fourth order if
one wishes to describe the same effects in a variational context.

Before going through the derivations, it is useful to
make contact with earlier work on flexoelectricity, 
by recalling the expansion of the force-constant matrix that was used in Ref.~\onlinecite{artlin},
\begin{eqnarray}
\tilde{\Phi}^{\bf q} &=& \tilde{\Phi}^{(0,\hat{\bf q})} -iq \tilde{\Phi}^{(1,\hat{\bf q})} 
- \frac{q^2}{2} \tilde{\Phi}^{(2,\hat{\bf q})} + \nonumber \\ 
  && i\frac{q^3}{3!} \tilde{\Phi}^{(3,\hat{\bf q})} + \frac{q^4}{4!} \tilde{\Phi}^{(4,\hat{\bf q})} + \ldots
\end{eqnarray}
The symbol $\hat{\bf q}$, appearing next to the perturbative order,
highlights that all the above expansion terms depend 
on the direction along which the differentiation is taken. 
(I stress that this dependence cannot 
be expressed in a tensorial form, as the macroscopic electric fields
contribution is nonanalytic in ${\bf q}$.) I shall drop this symbol
henceforth, keeping it implicit to avoid overburdening the notation.
We have, at a given order $n$,
\begin{equation}
D^{(n)}_{\kappa \alpha, \kappa' \beta} = \frac{(-i)^n}{n!}  \frac{1} {\sqrt{m_\kappa m_{\kappa'} }} 
  \tilde{\Phi}^{(n)}_{\kappa \alpha, \kappa' \beta}.
\end{equation}

\subsection{Order zero}

\label{sec:g0}

At the lowest order, the functional reads as
\begin{equation}
G^{(0)} = \langle v^{(0)} | \hat{D}^{(0)} | v^{(0)} \rangle.
\end{equation}
Since we are considering an acoustic phonon, we have 
\begin{equation}
v^{(0)}_{\kappa \alpha} = \hat{U}_\alpha \sqrt{ \frac{m_\kappa}{M}},
\label{v0}
\end{equation}
where $\hat{U}_\alpha$ is a real-space vector of unit length
and $M= \sum_\kappa m_\kappa$ is the total mass of the cell.
%
%
This clarifies the motivation for our choice of $M$ as the mass factor in
Eq.~(\ref{lqv}): via Eq.~(\ref{ufromv}) it is trivial to check that the
atomic sublattice displacements associated with $| v^{(0)} \rangle$ are
simply ${\bf u}^l_{\kappa} = \hat{\bf U}$, i.e. the amplitudes of the $| v(q) \rangle$
modes can be directly interpreted as a \emph{deformation} field in reciprocal space.

Because of translational invariance, of course, $G^{(0)}=0$.
%
There is, at first sight, a difficulty here as the ground state
at $q=0$ is threefold degenerate.
%
Such a degeneracy reflects the 
arbitrariness in choosing the acoustic phonon branch that one 
wishes to study (among one longitudinal and two transverse).
This is simply fixed by choosing a displacement direction $\hat{\bf U}$ 
to define the $q=0$ state via Eq.~(\ref{v0}) once and for all, and 
then sticking to it throughout the subsequent derivations;
such a procedure uniquely determines the higher-order $G^{(n)}$ 
expansion terms.

\subsection{Order two}

\label{sec:g2}

At second order, we have
\begin{eqnarray}
G^{(2)} &=& \langle v^{(1)} | \hat{D}^{(0)} | v^{(1)} \rangle + \nonumber \\
    && \langle v^{(1)} | \hat{D}^{(1)} | v^{(0)} \rangle + \langle v^{(0)} | \hat{D}^{(1)} | v^{(1)} \rangle + \nonumber \\
    && \langle v^{(0)} | \hat{D}^{(2)} | v^{(0)} \rangle.
    \label{eqg2}
\end{eqnarray}
By differentiating with respect to $\langle v^{(1)} |$, and by imposing 
that we are at a stationary point, we obtain the variational minimum 
condition for $| v^{(1)} \rangle$,
\begin{equation}
\hat{D}^{(0)} | v^{(1)}\rangle = - \hat{D}^{(1)} | v^{(0)} \rangle.
\label{eqv1}
\end{equation}
By replacing the dynamical matrix expansion terms
with their explicit expression in terms of the force-constant
matrix we obtain
\begin{equation}
\sum_{\kappa' \beta} \Phi^{(0)}_{\kappa \alpha, \kappa' \beta} \sqrt{\frac{ M }{m_{\kappa'}}} v^{(1)}_{\kappa' \beta}= 
   i \sum_{\kappa' \beta} \Phi^{(1)}_{\kappa \alpha, \kappa' \beta} \hat{U}_\beta,
\end{equation}
where we could remove the tilde on the $\Phi$ expansion terms after
observing that the crystal is not piezoelectric.
We obtain
\begin{equation}
v^{(1)}_{\kappa \alpha} = i \hat{U}_\beta \hat{q}_\gamma \, \sqrt{ \frac{m_\kappa}{M}}\, \Gamma^\kappa_{\alpha,\beta \gamma} ,
\end{equation}
where $\Gamma^\kappa_{\alpha,\beta \gamma}$
is the internal-strain response~\cite{artlin,Martin} of the cell, describing the displacement
of the atom $\kappa$ along $\alpha$ that is induced by a uniform strain of the type 
$\varepsilon_{\beta \gamma}$. 

By inserting Eq.~(\ref{eqv1}) into Eq.~(\ref{eqg2}) we can achieve a simpler
expression for the second-order functional,
\begin{equation}
G^{(2)} = \langle v^{(0)} | \hat{D}^{(2)} | v^{(0)} \rangle  - \langle v^{(1)} | \hat{D}^{(0)} | v^{(1)} \rangle .
\end{equation}
Finally, by replacing again $\hat{D}^{(n)}$ with $\Phi^{(n)}$, we have
\begin{equation}
G^{(2)} = \frac{1}{M} \hat{\bf U} \cdot \left[ -\frac{1}{2}  \Phi^{(2)} - \Gamma^{\rm T} \cdot \Phi^{(0)} \cdot \Gamma \right] \cdot \hat{\bf U}.
\label{eq41}
\end{equation}
It is straightforward to show~\cite{artlin,Born/Huang} that the above formula can be, in turn, rewritten as
\begin{equation}
G^{(2)} = \frac{ \mathcal{C}_{\alpha \lambda, \beta \gamma} \hat{q}_\lambda \hat{q}_\gamma  \hat{U}_\alpha \hat{U}_\beta }{ \rho_{\rm mass} },
\end{equation}
where $\bm{\mathcal{C}}$ is the relaxed-ion elastic tensor and $\rho_{\rm mass}$ is the mass density, 
thus recovering the well-known result of classical elasticity.
Note that the $\bm{\Gamma}$-dependent part in Eq.~(\ref{eq41}) is the internal-strain
relaxation contribution to the elastic constant, which is negative definite.
($\Phi^{(0)}$ has only positive or zero eigenvalues, given the requirement of lattice stability.)

Based on the considerations of Sec.~\ref{sec:discrete}, one can readily write the 
corresponding potential energy density (to be incorporated in the continuum Lagrangian density of
Sec.~\ref{sec:continuum}) as
\begin{equation}
E^{\rm elas} = \frac{\rho_{\rm mass}}{2} U^2 q^2 G^{(2)} =  \frac{1}{2} c_{\alpha \beta, i j}   U_\alpha U_\beta q_i q_j,
\label{eqelas}
\end{equation}
consistent with Eq.~(\ref{eq:lq}).
The fact that the elastic energy of Eq.~(\ref{eqelas}) enjoys an analytic expression
in a tensorial form rests on our assumption of a nonpiezoelectric crystal. As
we shall see in the following Sections, a careful consideration of electrostatic
long-range effects is necessary in order to achieve a closed expression at higher orders.

\subsection{Order four}

At the fourth order, the functional reads as
\begin{eqnarray}
G^{(4)} &=& \langle \tilde{v}^{(2)} | \hat{D}^{(0)} | \tilde{v}^{(2)} \rangle + \nonumber \\
    && \langle v^{(1)} | \hat{D}^{(1)} | \tilde{v}^{(2)} \rangle + \langle \tilde{v}^{(2)} | \hat{D}^{(1)} | v^{(1)} \rangle + \nonumber \\
    &&  \langle v^{(1)} | \hat{D}^{(2)} | v^{(1)} \rangle +  \nonumber \\
     &&  \langle \tilde{v}^{(2)} | \hat{D}^{(2)} | v^{(0)} \rangle +
        \langle v^{(0)} | \hat{D}^{(2)} | \tilde{v}^{(2)} \rangle + \nonumber \\
    &&    \langle v^{(1)} | \hat{D}^{(3)} | v^{(0)} \rangle + \langle v^{(0)} | \hat{D}^{(3)} | v^{(1)} \rangle + \nonumber \\ 
    &&   \langle v^{(0)} | \hat{D}^{(4)} | v^{(0)} \rangle,
    \label{g4_full}
\end{eqnarray}
to be minimized with the condition that $| \tilde{v}^{(2)} \rangle$ be orthogonal
to the subspace spanned by the three acoustic (A) branches at the zone center, i.e., 
$\langle \tilde{v}^{(2)} |v^{(0)}_{\rm A} \rangle = 0$. 
The tilde sign is meant to emphasize that $| \tilde{v}^{(2)} \rangle$, unlike $| v^{(0,1)} \rangle$, 
have a nonanalytic dependence on the wavevector direction $\hat{\bf q}$. This is due to the 
fact that the electrical boundary conditions are themselves a consequence of $\hat{\bf q}$:
the longitudinal component of the electric displacement field must vanish, whereas the electric 
field must vanish in the transversal plane. Thus, one should keep in mind that 
all tilded quantities are defined in ``mixed electrical boundary conditions''~\cite{Hong-11} (MEBC),
i.e. they implicitly contain the electrostatic contribution of the longitudinal fields
along the propagation direction.  If $\hat{\bf q}$ is fixed, as we have insofar assumed 
while performing the $q$-expansions, one does not really need to worry about this issue, whose detailed
treatment is deferred to Section~\ref{sec:electrostatics}.

Differentiation of $G^{(4)}$ with 
respect to $\langle \tilde{v}^{(2)} |$ leads to
\begin{equation}
\hat{D}^{(0)} | \tilde{v}^{(2)} \rangle = - \hat{Q} ( \hat{D}^{(1)} | v^{(1)} \rangle + \hat{D}^{(2)} | v^{(0)} \rangle ),
\label{tilv2}
\end{equation}
where the operator $\hat{Q}$ is a projector on the optical modes manifold.
By introducing the optical-phonon eigenmodes at the zone center, $|\tilde{v}^{(0)}_l \rangle$ (again, 
we use a tilde to remind the reader that these are eigenvectors of the dynamical matrix \emph{with the
electrostatic terms included}, i.e. they correspond to the correct longitudinal and transverse optical 
modes in the ${\bf q} \rightarrow 0$ limit), we can expand $| \tilde{v}^{(2)} \rangle$ as follows,
\begin{equation}
| \tilde{v}^{(2)} \rangle = -\sum_l  |\tilde{v}^{(0)}_l \rangle \frac{\tilde{f}_l}{M \tilde{\omega}_l^2},
\end{equation}
where we have introduced the \emph{energy flexocoupling coefficients} in MEBC 
along $\hat{\bf q}$,
\begin{equation}
\tilde{f}_l = M \langle \tilde{v}^{(0)}_l | ( \hat{D}^{(1)} | v^{(1)} \rangle + \hat{D}^{(2)} | \tilde{v}^{(0)} \rangle ).
\label{tildef}
\end{equation}
(A more in-depth discussion of these important quantities is deferred to Sec.~\ref{sec:fcoupling}.)
%
Based on this expression, we can simplify the fourth-order energy as
\begin{eqnarray}
G^{(4)} &=& -  \sum_l \frac{\tilde{f}_l^2}{M^2 \tilde{\omega}_l^2} +\nonumber \\
 && \langle v^{(1)} | \hat{D}^{(2)} | v^{(1)} \rangle + \nonumber \\
 && \langle v^{(1)} | \hat{D}^{(3)} | v^{(0)} \rangle + \langle v^{(0)} | \hat{D}^{(3)} | v^{(1)} \rangle +  \nonumber \\
 && \langle v^{(0)} | \hat{D}^{(4)} | v^{(0)} \rangle.
\label{g4}
\end{eqnarray} 
Remarkably, just like the flexoelectric tensor, one can decompose the 
strain-gradient contribution to the acoustic frequency dispersion
into three parts.
The first line in Eq.~(\ref{g4}) describes the contribution of
\emph{lattice-mediated} effects, i.e. is related to the (adiabatic) relaxation of the optical modes
(internal strains) within the deformation field produced by the acoustic phonon. 
The second and third line is a ``\emph{mixed}'' (lattice and electronic) contribution, 
due to the dispersion of the (nonpolar) optical modes that couple directly
to the strain, and is absent in materials like SrTiO$_3$ (see Section~\ref{sec:results}).
The fourth line is the \emph{purely electronic} (``frozen-ion'') contribution; it 
is sometimes referred to as the ``self-dispersion'' of the acoustic branch,
and is always present even in the simplest monoatomic model. 

The functional $G^{(4)}$ can be readily interpreted as the
hyperelastic coefficient in MEBC, referred to the propagation direction $\hat{\bf q}$
and to the polarization $\hat{\bf U}$ of the branch,
\begin{equation}
G^{(4)} = 
\frac{\tilde{h}}{\rho_{\rm mass}}.
\end{equation}
One would be tempted, at this point, to establish a direct link between the 
coefficients $\tilde{h}$ 
and the ${\bf h}$-tensor, similarly to what we have done in Sec.~\ref{sec:g2} 
for the classical elasticity case at $\mathcal{O}(q^2)$. 
Before doing this, however, we need to stop for a second and
deal with the electrostatic energy. This, as we said, is implicitly 
contained in the $\hat{D}^{(n)}$ operators, which implies that 
$\tilde{h}$ contains \emph{both} the SGE and Maxwell energy 
(see Sec.~\ref{sec:continuum}). 
%
We need to separate the two in order to achieve a proper 
tensorial representation as that of Eq.~(\ref{eq:lq}).

\subsection{Electrostatic energy}

\label{sec:electrostatics}

%
%
The formulas derived insofar work equally well for a longitudinal or 
transversal phonon, but one must keep in mind that the electrical 
boundary conditions, hard-wired in the definition of the 
$|\tilde{v}^{(2)}\rangle$ eigenmodes, differ depending on the wavevector 
direction and on the transverse versus longitudinal regime. 
For this reason, this theory cannot be directly transformed into
an energy functional of the system. Before taking such a step,
one needs to separate the electrostatics from the other interactions,
and describe them explicitly in a physically consistent form.

The macroscopic fields concern each of the expansion terms, $D^{(0)}$, $D^{(1)}$ and $D^{(2)}$, 
whose behavior is nonanalytic.~\cite{artlin,rmm_thesis}
Such a direction dependence is famously responsible, in the case of optical
phonons, for the LO-TO splitting at the zone center. 
In the acoustic case under consideration here the eigenvectors at lower ($n=0,1$) orders are not affected 
(at order $n=0$ this is an obvious consequence of the acoustic sum rule; at $n=1$ this follows our 
assumption that the crystal is nonpiezoelectric). 
It is then convenient, first of all, to rewrite the $G^{(4)}$ functional
of Eq.~(\ref{g4_full}) by eliminating its explicit dependence on 
$| \tilde{v}^{(2)} \rangle$.
To this end, we combine Eq.~(\ref{eqv1}) and Eq.~(\ref{tilv2}) to
write
$$
| \tilde{v}^{(2)} \rangle = - \widetilde{D}^{(0)} \,
 \bar{D}^{(2)} \, | v^{(0)} \rangle,
$$
where 
\begin{equation}
\bar{D}^{(2)} = D^{(2)} - ( D^{(1)} \cdot \widetilde{D}^{(0)} \cdot D^{(1)} ),
\end{equation}
and $\widetilde{D}^{(0)}$ is the \emph{pseudoinverse}~\footnote{The
  matrix inversion is performed only on the optical modes subspace, leaving
  a null eigenvalue on the translational part.}
of the zone-center
dynamical matrix. (We have dropped the hat symbols starting from this Section, as it
should be clear by now that the ${D}^{(n)}$ represent Hermitian operators.)
After substituting $| \tilde{v}^{(2)} \rangle$ in Eq.~(\ref{g4_full}), we 
obtain
\begin{eqnarray}
G^{(4)} &=& -\langle v^{(0)} | \, 
  \bar{D}^{(2)} \,
 \widetilde{D}^{(0)} \,
 \bar{D}^{(2)} \, | v^{(0)} \rangle \nonumber \\ 
 && + \langle v^{(1)} | D^{(2)} | v^{(1)} \rangle \\
 && + \langle v^{(1)} | D^{(3)} | v^{(0)} \rangle + \langle v^{(0)} | D^{(3)} | v^{(1)} \rangle \nonumber \\ 
 && + \langle v^{(0)} | D^{(4)} | v^{(0)} \rangle,
\end{eqnarray}
We have thus achieved an expression for $G^{(4)}$ where the nonanalyticity
is only carried by the operators, and not by the eigenvectors.

We shall proceed by separating such nonanalytic (NA) multipolar interactions from the 
dynamical matrix, i.e. write
\begin{equation}
D^{(n)} = \mathcal{D}^{(n)} + D^{(n),{\rm NA}},
\end{equation}
where $\mathcal{D}^{(n)}$ represents the expansion terms of the 
dynamical matrix \emph{without} macroscopic fields, and 
$D^{(n),{\rm NA}}$ are the analogous expansion terms of
\begin{equation}
 \langle \alpha \kappa | D^{{\bf q},{\rm NA}} | \beta \kappa' \rangle = 
  \frac{1}{ \sqrt{m_\kappa m_{\kappa'} } } \Phi^{{\bf q},{\rm NA}}_{\alpha \kappa, \beta \kappa'}.
\end{equation}  
The explicit expression of $\Phi^{{\bf q},{\rm NA}}$, as derived in
Ref.~\onlinecite{artlin}, consists in the electrostatic interaction between
the multipoles induced by atomic displacements. This can be expressed in
the present context as
\begin{equation}
D^{{\bf q},{\rm NA}} = \frac{4 \pi }{\Omega M} \frac{ | \mathcal{Q}{(q)} \rangle \langle \mathcal{Q}{(q)} | } { \xi{(q)} },
\end{equation}
where
\begin{equation}
\mathcal{Q}^{(q)}_{\kappa \beta} = \sqrt{ \frac{M}{m_\kappa} } \left( -iq Q^{(1,\hat{\bf q})}_{\kappa \beta}
   - \frac{q^2}{2} Q^{(2,\hat{\bf q})}_{\kappa \beta} + i\frac{q^3}{3!} Q^{(3,\hat{\bf q})}_{\kappa \beta} + \ldots \right),
\end{equation}   
and
\begin{equation}
\xi{(q)} = q^2 \epsilon^{(2,\hat{\bf q})}_\infty + q^4 \epsilon^{(4,\hat{\bf q})}_\infty + \ldots
\end{equation}
Here $Q^{(n,\hat{\bf q})}_{\kappa \beta}$, 
are the longitudinal (along $\hat{\bf q}$) components of the dynamical
multipole tensors associated to the displacement of an atom $\kappa$ along $\beta$; 
for example,
\begin{equation}
Q^{(1,\hat{\bf q})}_{\kappa \beta} = Z^*_{\kappa,\alpha \beta} \hat{q}_\alpha
\end{equation}
is the longitudinal component of the dynamical dipole tensor $Z^*_{\kappa,\alpha \beta}$,
more commonly known as the Born effective charge tensor.
$\epsilon_\infty^{(2,\hat{\bf q})}$ is the corresponding
element of the electronic (high-frequency) dielectric tensor, 
\begin{equation}
\epsilon_\infty^{(2,\hat{\bf q})} = \hat{\bf q} \cdot \bm{\epsilon}_\infty \cdot \hat{\bf q},
\end{equation}
and 
$\epsilon^{(4,\hat{\bf q})}_\infty$ is related to the (purely electronic) dielectric 
dispersion. (The latter quantity is irrelevant in the context of the present work,
and we won't discuss it any further.)
At the lowest (zero) order we have the usual~\cite{rmm_thesis,Cochran/Cowley} dipole-dipole term, which is responsible
for the LO-TO splitting in polar crystals, 
\begin{equation}
D^{(0,{\rm NA})} = \frac{4 \pi }{\Omega M} 
  \frac{ | Z(\hat{\bf q}) \rangle \langle Z(\hat{\bf q}) | } { \hat{\bf q} \cdot \bm{\epsilon}_\infty \cdot \hat{\bf q} },
\end{equation}
where 
\begin{equation}
Z_{\kappa \beta}^{(\hat{\bf q})} = \sqrt{ \frac{M}{m_\kappa} }  Z^*_{\kappa,\alpha \beta} \hat{q}_\alpha,
\end{equation}
while at higher orders in $q$ 
quadrupoles, octupoles and higher-order multipoles are also involved.
  
After rewriting the pseudoinverse of the zone-center dynamical matrix 
  by means of the Sherman-Morrison formula,
  \begin{equation}
  \widetilde{D}^{(0)} = \widetilde{\mathcal{D}}^{(0)} - \frac{4 \pi} {M \Omega} 
 \frac{\widetilde{\mathcal{D}}^{(0)}  | Z (\hat{\bf q})\rangle \langle Z(\hat{\bf q}) |\widetilde{\mathcal{D}}^{(0)} }
 {{\hat{\bf q}} \cdot \bm{\epsilon}_{\rm static} \cdot {\hat{\bf q}}}.
\end{equation}
some cumbersome but otherwise straightforward algebra leads to the following result for the fourth-order energy,
\begin{eqnarray}
G^{(4)} &=& -\langle v^{(0)} | \, 
  \bar{\mathcal{D}}^{(2)} \,
 \widetilde{\mathcal{D}}^{(0)} \,
 \bar{\mathcal{D}}^{(2)} \, | v^{(0)} \rangle \nonumber \\ 
 && + \langle v^{(1)} | \mathcal{D}^{(2)} | v^{(1)} \rangle \\
 && + \langle v^{(1)} | \mathcal{D}^{(3)} | v^{(0)} \rangle + \langle v^{(0)} | \mathcal{D}^{(3)} | v^{(1)} \rangle \nonumber \\ 
 && + \langle v^{(0)} | \mathcal{D}^{(4)} | v^{(0)} \rangle \nonumber \\
 && + \frac{4 \pi \Omega}{M} \frac{ \mu_{\hat{\bf q}}^2 }{{\hat{\bf q}} \cdot \bm{\epsilon}_{\rm static} \cdot {\hat{\bf q}}}.
\label{g4sep}
\end{eqnarray}
Here $\mu_{\hat{\bf q}}$ is the longitudinal (along $\hat{\bf q}$) component of the 
total (electronic and ionic) flexoelectric polarization induced by the 
strain gradient that is associated to the phonon eigenmode,
\begin{equation}
\mu_{\hat{\bf q}} = -\hat{q}_\alpha \hat{U}_\beta \hat{q}_\gamma \hat{q}_\lambda \mu^{\rm I}_{\alpha \beta, \gamma \lambda}, 
\end{equation}
where 
\begin{equation}
\mu^{\rm I}_{\alpha \beta, \gamma \lambda} = \frac{ d P_\alpha}{d \eta_{\beta,\gamma \lambda} }
\end{equation}
is the type-I~\cite{artlin} flexoelectric tensor [$\eta_{\beta,\gamma \lambda}$ is the 
mixed partial derivative along $r_\gamma$ and $r_\lambda$ of the displacement field $u_\beta ({\bf r})$].


The above derivation has led to a simple and physically transparent result:
the nonanalytic contribution to $G^{(4)}$, 
\begin{equation}
G^{(4,{\rm NA})} = \frac{4 \pi \Omega}{M} \frac{ \mu^2 }{\epsilon_{\rm static}},
\end{equation}
simply corresponds to the Maxwell energy density of the flexoelectrically induced electric fields,
%
\begin{equation}
E^{\rm Max} =   \frac{\rho_{\rm mass}}{2} U^2 q^4 G^{(4,{\rm NA})},
\end{equation}
as we anticipated in Sec.~\ref{sec:continuum}.
The remainder of $G^{(4)}$ is analytic, i.e. it can be expressed in
a tensorial form, and can be directly associated with the
strain-gradient elasticity term of Eq.~(\ref{eq:lq}), 
\begin{equation}
E^{\rm SGE} =  
  \frac{\rho_{\rm mass}}{2} U^2 q^4 \left( G^{(4)}  - G^{(4,{\rm NA})}  \right).
\label{sgefromg}  
\end{equation}
Thus, the above derivation provides us with a comforting
proof that our fourth-order energy functional is indeed correct,
and physically consistent with the continuum formulation of
Sec.~\ref{sec:continuum}.

Before moving on, it is useful to emphasize two further facts
regarding the connection between the lattice-dynamical result of
Eq.~(\ref{g4sep}) and the continuum functional of Eq.~(\ref{eq:lq}).
First, the decomposition of the dynamical matrix into analytic and 
nonanalytic contributions is nonunique, which relates to the arbitrariness,
discussed in Sec.~\ref{sec:recip}, in the separation 
between flexo-electrostatic and strain-gradient elasticity 
contributions to the energy. As we said, this can
be readily interpreted as a \emph{gauge freedom} of the theory.
%
%
%
%
%
(In Sec.~\ref{sec:results} we shall quantitatively
assess how different choices of the reference potential affect the partition between 
$E^{\rm SGE}$ and $E^{\rm Max}$ in some selected cases.)
Second, it should be noted that each of
the different contributions (lattice-mediated, mixed and
electronic) to $E^{\rm SGE}$ [as inferred from Eq.~(\ref{g4sep}) and 
Eq.~(\ref{sgefromg})] enjoys a slightly different tensorial representation. 
Leaving aside the mixed term (which in any case is absent from the
calculations presented in Sec.~\ref{sec:results}),
the electronic term can be directly mapped into a symmetrized
form,
\begin{eqnarray}
E^{\rm SGE,el} &=&  \frac{1}{2} U_\alpha U_\beta q_i q_j q_k   q_l \bar{h}_{\alpha \beta, ijkl}, \\
\bar{h}_{\alpha \beta, ijkl} &=& \frac{1}{4!} \sum_{\kappa \kappa'} \Phi^{(4,ijkl)}_{\kappa \alpha, \kappa' \beta},
\label{hbar}
\end{eqnarray} 
where the bar symbol is a reminder that lattice-mediated
effects are not included.
The short-circuit $\bar{h}$ coefficient along a given direction is related to 
the corresponding MEBC coefficient, $\tilde{h}^{\rm el} = \langle v^{(0)} | D^{(4)} | v^{(0)} \rangle$,
by
\begin{equation}
\tilde{h}^{\rm el} = \bar{h} + 
  4 \pi  \frac{ \bar{\mu}_{\hat{\bf q}}^2 }{\epsilon_{\infty}},
  \label{hbarsc} 
\end{equation}  
where the second term on the right-hand side is the electrostatic 
energy due to the purely electronic flexoelectric effect. (Note that
the longitudinal flexoelectric coefficient $\bar{\mu}_{\hat{\bf q}}$
can be inferred from the dynamical octupole tensor.~\cite{Resta-10,Hong-11,artlin})
As we shall demonstrate shortly, the lattice-mediated contribution is most naturally 
written, instead, in a \emph{separable} type-II representation, 
\begin{eqnarray}
E^{\rm SGE,LM} &=& \frac{1}{2} \nabla \bm{\varepsilon} \cdot {\bf H}^{\rm LM} \cdot \nabla \bm{\varepsilon}. 
\end{eqnarray}
%

\subsection{Energy flexocoupling tensor}

\label{sec:fcoupling}

Assuming that we have suppressed the macroscopic electric
fields (after associating them with a given energy 
reference that we choose once and for all), the strain-gradient 
elastic energy associated with the deformation field reads as
\begin{eqnarray}
G^{(4,{\rm SGE})} 
 &=& -  \sum_l \frac{f_l^2}{M^2 \omega_l^2} +\nonumber \\
 && \langle v^{(1)} | \mathcal{D}^{(2)} | v^{(1)} \rangle + \nonumber \\
 && \langle v^{(1)} | \mathcal{D}^{(3)} | v^{(0)} \rangle + \langle v^{(0)} | \hat{D}^{(3)} | v^{(1)} \rangle +  \nonumber \\
 && \langle v^{(0)} | \mathcal{D}^{(4)} | v^{(0)} \rangle.
\end{eqnarray}
Here we have introduced the \emph{short-circuit} energy flexocoupling coefficients,
\begin{equation}
f_l = M \langle v^{(0)}_l | ( \mathcal{D}^{(1)} | v^{(1)} \rangle + \mathcal{D}^{(2)} | v^{(0)} \rangle ),
\end{equation}
which describe the coupling between an arbitrary strain gradient component and the
\emph{transverse} optical (TO) modes at $\Gamma$; consistently, $\omega_l$ now stands
for the frequency of the $l$-th TO mode.
It is convenient to express the dependence on $\hat{\bf q}$ and ${\bf U}$
explicitly, which leads to a (type-I) tensor representation for the $f_l$ 
coefficients,
\begin{equation}
f_l = -\hat{U}_\beta \hat{q}_\gamma \hat{q}_\lambda f^{\rm I}_{l \beta,\gamma \lambda}.
\end{equation}
Just like for the flexoelectric tensor, one can readily switch back and forth
from a type-I to a type-II representation~\cite{artlin} (recall that the
former is associated to second gradients of the displacement, while the
latter is associated to first gradients of the symmetric strain) of the 
flexocoupling tensor via
\begin{equation}
 f^{\rm I}_{l \beta,\gamma \lambda} = {\rm sym}_{(\gamma \lambda)} \, f^{\rm II}_{l \lambda, \beta \gamma}
\end{equation}
 This allows us to write the lattice-mediated contribution to the SGE
energy directly in a separable type-II form, as required by Eq.~(\ref{eqsge}),
\begin{eqnarray}
H^{\rm LM}_{\beta \beta', \gamma \gamma' \lambda \lambda'} &=& -\frac{1}{M \Omega} 
   \sum_l  \frac{ f^{\rm II}_{l \beta,\gamma \lambda}  f^{\rm II}_{l \beta',\gamma' \lambda'}}{\omega_l^2}.
   \label{hlm}
\end{eqnarray}  
This also shows that the lattice-mediated contribution is 
always negative, as expected.

%

The ${\bf f}$-tensor introduced here bear a close resemblance to the flexocoupling 
coefficients described, e.g., by Yudin and Tagantsev~\cite{Yudin-13} (YT), with the important difference 
that the former have the physical dimension of an energy, while the latter are expressed as
a voltage.
In a simple cubic material we can trace an exact link between 
the two by writing 
\begin{equation}
f^{\rm YT}_{l} = \frac{f_{l}} {Z^*_{l}},
\label{voltage}
\end{equation}
i.e. by dividing the energy coefficient by the dynamical charge associated to 
the mode $l$. 
%
%
%
Based on such arguments, one could be tempted to rewrite our expressions for the 
strain-gradient energy by using the voltage coefficients as defined in Eq.~(\ref{voltage}).
This, however, would only be applicable to a very restricted range
of materials: 
First, the energy coefficients (unlike the voltage ones) can be used to
describe the coupling between a strain gradient and a \emph{nonpolar} optical mode 
-- these, of course, do not contribute to the polarization, but they do contribute to the 
energetics (we shall see a concrete example in Sec.~\ref{sec:results}). 
Second, the mode effective charge appearing at the denominator in Eq.~(\ref{voltage}) is 
generally a three-dimensional vector, not a scalar -- such a formula can only be effectively 
applied to cubic crystals, while its adaptation to less symmetric material classes remains 
unclear.
%
%
Clearly, our present formalism based on the energy coefficients $f_l$ is more 
general without entailing any additional burden in the formulas, and therefore 
preferrable.

We can now use the above derivations to connect to earlier \emph{ab initio} works on 
flexoelectricity. For example, one can express the $\mathcal{O}(q^2)$ contribution to the 
acoustic eigenmode (under short circuit EBC), $|v^{(2)} \rangle$, in two different tensorial 
forms: either based on $f_{l \beta,\gamma \lambda}$,
\begin{eqnarray}
| v^{(2)} \rangle &=& -\sum_l  |v^{(0)}_l \rangle \frac{f_l}{M \omega_l^2} \nonumber \\
  &=& -\sum_{l}  |v^{(0)}_{l} \rangle 
       \frac{U_\beta \hat{q}_\gamma \hat{q}_\lambda f_{l \beta,\gamma \lambda}}{M \omega_{l}^2},
\label{v2a}
\end{eqnarray}
or in terms of the \emph{flexoelectric internal-strain tensor}, 
$L^\kappa_{\alpha \lambda, \beta \gamma}$, that was introduced in
Ref.~\onlinecite{artlin},
\begin{equation}
v^{(2)}_{\kappa \alpha} = - \hat{U}_\beta \hat{q}_\gamma \hat{q}_\lambda \, 
         \sqrt{ \frac{m_\kappa}{M}}\, L^\kappa_{\alpha \lambda, \beta \gamma }.
\label{v2b}
\end{equation}
[By comparing Eq.~(\ref{v2a}) and Eq.~(\ref{v2b}) one trivially obtains ${\bf L}$
as a function of $f_{l \beta, \gamma \lambda}$ and the optical mode 
eigendisplacements and frequencies.] 

It is useful in this context to express the lattice-mediated (LM) contribution
to the flexoelectric tensor as
\begin{equation}
\mu^{\rm I,LM}_{\xi \beta, \gamma \lambda} = \frac{1}{\Omega} \sum_l  \frac{Z^*_{l \xi} 
  f_{l \beta, \gamma \lambda}}{M \omega_{l}^2} 
\end{equation}
where we have introduced the dynamical charge associated to the $l$-th polar mode, 
\begin{equation}
Z^*_{l \alpha} = \sum_{\kappa \rho} Z^*_{\kappa,\alpha \rho} \sqrt{\frac{M}{m_\kappa}} \langle \kappa \rho  |  v^{(0)}_l \rangle.
\end{equation}
(Usually the mass factor $M$ is assumed to be arbitrary; for the above formulas to be valid, it is 
necessary to choose it as the total mass of the unit cell.)
The above formulas nicely parallel the known expression for the
lattice contribution to the dielectric permittivity, which 
in the present notation reads as
\begin{equation}
\epsilon^{\rm ion}_{\alpha \beta} = \sum_l \frac{4 \pi}{M \Omega} \frac{Z^*_{l \alpha} Z^*_{l \beta}}{\omega_l^2}.
\end{equation}
As we shall see shortly, the presence of $\omega_l^2$ at the denominator
in the expressions for the hyperelastic (SGE) energy, flexoelectric polarization
and dielectric permittivity has important implications in materials like SrTiO$_3$:
these are characterized by a ``soft'' polar mode with small frequency, which means 
that its contributions to the above physical quantities can be very large.
%
%
%
%
%

\subsection{Special case: cubic perovskites}


Since the $\bm{\Gamma}$-tensor (referring to the internal atomic
relaxations induced by a uniform strain) identically vanishes in the cubic
perovskite structure,
the expression for the fourth-order functional 
simplifies to
\begin{eqnarray}
G^{(4,{\rm SGE})} &=& -  \sum_l \frac{f_l^2}{M^2 \omega_l^2} + \langle v^{(0)} | \mathcal{D}^{(4)} | v^{(0)} \rangle.
\end{eqnarray} 
In other words, the ``mixed'' contribution to the SGE energy vanishes,
leaving only the electronic and lattice-mediated terms behind.
Thanks to the symmetry, the fifteen normal modes of the crystal can then be grouped together as
five vector fields, by breaking up the index $l=1,\ldots,15$ into a mode index $j=1,\ldots,5$ 
and a Cartesian index $\alpha$. Correspondingly, the flexocoupling tensor can be written in
a form that more closely resembles that of the flexoelectric tensor,
\begin{equation}
f_{l\beta,\gamma \lambda} = f^j_{\alpha \beta,\gamma \lambda}.
\end{equation}
This form is particularly convenient, as for a given $j$ the tensor
$f^j_{\alpha \beta,\gamma \lambda}$ has the same symmetries as the flexoelectric 
tensor, e.g., in cubic materials there are only three independent components.

Most importantly, in incipient ferroelectrics like SrTiO$_3$ the lowest polar mode
has a small frequency, and is therefore expected to dominate the energetics (given
the $\omega^{-2}$ prefactor in $E^{\rm SGE}$), provided that the flexocoupling 
coefficients $f^j_{\alpha \beta,\gamma \lambda}$ are all comparable in magnitude.
Under such conditions one can, therefore, neglect the contributions from the stiff 
polar modes, and retain only the soft mode, with frequency $\omega_1$, that we describe 
as a three-dimensional vector. 
In order to avoid overburdening of
the indices, we can choose a specific propagation $(\hat{\bf q})$ and displacement
$(\hat{\bf U})$ direction. 
The relevant components of the flexocoupling tensor can be then represented by
a vector quantity, ${\bf f}_1$, where the subscript refers to the lowest TO1
mode, and is related to the full tensor as
\begin{equation}
f_{1\alpha} = f^{j=1}_{\alpha \beta,\gamma \lambda} \hat{U}_\beta \hat{q}_\gamma \hat{q}_\lambda.
\end{equation}
%
One can then perform the following approximations
\begin{eqnarray}
\frac{E^{\rm SGE}}{U^2 q^4} &\approx& - \frac{1}{2 M \Omega} \frac{ |{\bf f}_1|^2}{\omega_1^2}, \\
\mu_{\hat{\bf q}} &\approx& \frac{1}{\Omega} \frac{ Z^*_1  \, (\hat{\bf q} \cdot {\bf f}_1)}{M \omega_1^2}, \\
\epsilon_{\rm static}  &\approx& \frac{4 \pi}{M \Omega} \frac{ (Z^*_1)^2 }{\omega_1^2}.
\end{eqnarray}
Based on the above, we readily obtain the dominant contribution to the electrostatic 
energy,
\begin{equation}
\frac{E^{\rm Max}}{U^2 q^4} = \frac{4\pi}{2} \frac{\mu_{\hat{\bf q}}^2}{\epsilon_{\rm static}} \approx 
\frac{1}{2 M \Omega} \frac{ (\hat{\bf q} \cdot {\bf f}_1)^2 }{\omega_1^2}.
\label{enel}
\end{equation}
Summarizing, the overall strain gradient-related contributions to the total energy  
go like
\begin{equation}
\frac{E^{\rm tot}}{U^2 q^4} \approx \frac{1}{2 M \Omega} \frac{(\hat{\bf q} \cdot {\bf f}_1)^2 - |{\bf f}_1|^2}{\omega_1^2}.
\end{equation}
This means that the soft-mode contribution is irrelevant along the longitudinal
direction, but can be large for phonons that produce a transverse flexoelectric
polarization, where it may lead to a considerable softening of the elastic response at
short length scales. (Such a length scale, in fact, diverges as $\omega_1 \rightarrow 0$.)
This is fully consistent with the observation of Refs.~\onlinecite{Axe-70} and 
\onlinecite{Kvasov-15} that
the dominant source of dispersive behavior in the acoustic phonon branch is due to the
interaction with a low-energy optical mode; LO modes lie higher
in energy, and therefore contribute comparatively less to the 
anomalous acoustic dispersion described in the above works.

\subsection{Experimental determination of ${\bf f}$}

\label{sec:g6}

Based on the conclusion of the previous Section, that the dispersion of transversal acoustic 
(TA) modes is dominated by their interaction with the soft polar branch, 
Kvasov and Tagantsev~\cite{Kvasov-15} proposed that the experimentally measured
phonon frequencies may be used to infer the value of the corresponding flexocoupling tensor components, 
$f^1_{\alpha \beta, \gamma \lambda}$.
(Our numerical results of Section~\ref{sec:results} provide quantitative support to this 
statement.)
The authors correctly observed that the values of the coefficients determined this way are 
inherently dynamic quantities (i.e., directly depend on the atomic masses). 
This fully agrees with the conclusions of this work: one can easily show that
$f^j_{\alpha \beta, \gamma \lambda}$ as defined here coincide with Eqs.~(42) and~(43) of 
Ref.~\onlinecite{artlin}, where the mass dependence is explicit.

A related question that has been raised recently consists in whether or not 
two separately measurable contributions to ${\bf f}$ exist, one of static and the 
other of dynamic nature.
Ref.~\onlinecite{Kvasov-15} claims that the answer is positive: the dynamic and static 
effects would manifest themselves differently once the expansion of the TA frequency 
is pushed to higher orders in the wavevector $q$, allowing in principle for an experimental
separation of the two.

By using the theoretical formalism developed in this work, it is not difficult to verify 
this statement -- it suffices to apply the $2n+1$ theorem to higher perturbatives orders 
in $q$, and look for any signature of the ``flexodynamic'' tensor introduced in 
Ref.~\onlinecite{Kvasov-15}.
Specializing to the case of cubic SrTiO$_3$, the sixth-order functional reads as
\begin{eqnarray}
G^{(6)} &=& \langle v^{(2)} | (\hat{D}^{(2)} - X^{(2)}) | v^{(2)} \rangle + \nonumber \\
    &&  \langle v^{(2)} | \hat{D}^{(4)} | v^{(0)} \rangle +
        \langle v^{(0)} | \hat{D}^{(4)} | v^{(2)} \rangle + \nonumber \\
    &&   \langle v^{(0)} | \hat{D}^{(6)} | v^{(0)} \rangle,
    \label{eqg6}
\end{eqnarray}
where we have used the fact that the phonon eigenmode contains only 
even-order contributions (i.e. $| v^{(1,3,\ldots)} \rangle = 0$).
The above expression, as $G^{(4)}$, only depends on the 
flexocoupling coefficients $f_l$ via $| v^{(2)} \rangle$, i.e.
there is no direct dependence on the ``flexodynamic'' effect, 
contrary to the arguments of Ref.~\onlinecite{Kvasov-15}. 
In more detail, for a TA mode the dominant term at low temperatures is the 
first row of Eq.~(\ref{eqg6}), which can be written as
\begin{eqnarray}
G^{(6)} &\approx& \sum_{jl}  \frac{f_l f_j}{M^2 \omega_l^2 \omega_j^2} 
  \left( \langle v^{(0)}_l | \hat{D}^{(2)} | v^{(0)}_j \rangle - \frac{\mathcal{C}}{ \rho_{\rm mass}  } \delta_{lj} \right) \nonumber \\
 &\approx& \frac{ {f}_1^2 ( g_{11} - \mathcal{C} \Omega ) }{ M^3 \omega_1^4 }.
\label{eqg6a}
\end{eqnarray}  
Here $\mathcal{C}$ is the relevant component of the elastic tensor, $M$ and $\Omega$ are
as usual the total mass and volume of the primitive cell, and we have introduced,
in analogy with the definition of the energy flexocoupling coefficient $f_l$, the
\emph{correlation matrix}~\cite{Kvasov-15,Yudin-13}
\begin{equation}
g_{lj} = M \langle v^{(0)}_l | \hat{D}^{(2)} | v^{(0)}_j \rangle.
\end{equation}
$g_{lj}$ has the dimension of energy; it describes the quadratic dispersion of 
the optical branches and their mutual interaction at $\mathcal{O}(q^2)$.
%
%
The discrepancy between our conclusions and those of Ref.~\onlinecite{Kvasov-15}
may originate from the inclusion of a kinetic cross-term between the strain and polar degrees of
freedom in the phenomenological thermodynamic functional of Refs.~\onlinecite{Kvasov-15} and
\onlinecite{Yudin-13}; such a term is absent from our lattice-dynamical treatment, which 
is based on a normal mode representation.

This derivation corroborates the argument of Ref.~\onlinecite{artlin}: 
distinguishing between dynamic and static contributions to the flexoelectric
effect is somewhat artificial, as the two quantities are not separately measurable.
We stress that, even if the individual components of the flexoelectric tensor are
inherently dynamic quantities, and therefore relevant to sound waves, they are 
perfectly appropriate to address static phenomena as well,~\cite{artlin} 
thus there is no need to consider a different tensor for each context.

\subsection{Static or dynamic?}

\label{sec:static}

In the previous Section we have questioned the dynamic or static 
nature of some key quantities involved in the present formalism, i.e., 
the flexocoupling coefficients. 
This is a natural context to raise the same question about the SGE tensor
components: Are they static or a dynamic?
To answer this question, one needs to go
back to the formulas we have derived so far, and inspect them to
see whether they contain any explicit dependence on the atomic masses:
if they do, then the corresponding physical quantity must be a 
dynamic one. 
%

We shall separately focus on two physical quantities, the purely electronic and 
lattice-mediated contributions to the SGE energy, as described respectively by
the tensors $\bar{\bf h}$ of Eq.~(\ref{hbar}) and ${\bf H}^{\rm LM}$
of Eq.~(\ref{hlm}). 
%
Clearly, the electronic tensor $\bar{\bf h}$ is a static one: It
is independent of the masses [it can be written as a double sublattice sum of
the force-constant matrix at fourth order in $q$, see Eq.~(\ref{hbar})], consistent
with its physical interpretation. (One can think, at least in the
context of a calculation, of forcing the atoms by
hand into a macroscopic strain-gradient pattern, and let the electrons relax
in such a static deformation field.)
The lattice-mediated part, on the other hand, is \emph{generally} dynamic in nature,
consistent with the known~\cite{artlin} mass dependence of the flexoelectrically induced
internal strains. 
To see this, it is instructive to write ${\bf H}^{\rm LM}$ in terms of
zone-center force-constant matrix and the internal strain response tensor, ${\bf L}$,
\begin{equation}
{\bf H}^{\rm LM} = -{\bf L} \cdot \Phi^{(0)} \cdot {\bf L},
\end{equation}
which follows trivially from Eq.~(\ref{v2b}) after observing that
$H^{\rm LM} = - \rho_{\rm mass} \langle v^{(2)} | \mathcal{D}^{(0)}| v^{(2)} \rangle$.
The individual components of ${\bf L}$ are dynamic,~\cite{artlin} and this characteristic
directly propagates to ${\bf H}^{\rm LM}$. 
%

The latter observation does not imply by any means that the scopes of the present 
theory are limited to dynamic effects: In fact, the present definition of ${\bf H}^{\rm LM}$ 
is perfectly suited to describing the energy associated with \emph{static} deformation fields as well.
To see this, suppose we have an inhomogeneous deformation field at rest under the
action of a static external load (e.g., applied to a far-away portion of the
crystal).
Then, due to the mechanical equilibrium condition, the mass dependence disappears~\cite{artlin} 
from the \emph{effective} internal strains that arise at any point in the
crystal and, consequently, from the overall SGE energy.
Thus, the same considerations that have been made in the case of 
flexoelectricity are equally valid in the case of strain gradient elasticity:
individual tensor components are dynamic, but their overall contribution
becomes static (and hence, mass-independent) at mechanical rest.

\section{Results: Bulk {S\lowercase{r}T\lowercase{i}O$_3$}}

\label{sec:results}

\subsection{Computational parameters}

\begin{table}
\begin{center}
\begin{ruledtabular}
\begin{tabular}{crrr rrr}
 & TO1  & TO2  & TO3 & LO1 & LO2 & LO3 \\
\hline 
 $\omega_l$ (cm$^{-1}$) &   36.33  & 170.18  & 556.20  & 164.32      & 457.46 & 790.77 \\
 $Z^*_l$ ($e$)         &     22.65  &   5.97  &  11.64  &   0.41      &   8.05 & 24.88                         
\end{tabular}
\end{ruledtabular}
\end{center}
\caption{ 
Lattice-dynamical properties of bulk SrTiO$_3$. The table shows the frequency
and dynamical charge of the IR-active zone-center optical modes. 
(The silent mode has a frequency of $\omega_{\rm S}=234.0$ cm$^{-1}$ and
its dynamical charge is zero by symmetry.)
The calculated dielectric constants are $\epsilon_\infty$=6.18, $\epsilon_{\rm static}$=1846.0. 
Calculations are performed at the theoretical equilibrium lattice parameter $a_0$=3.85 \AA. \label{tab1}}
\end{table}

\begin{figure*}
\begin{center}
\begin{tabular}{rr}
\includegraphics[width=2.8in,clip]{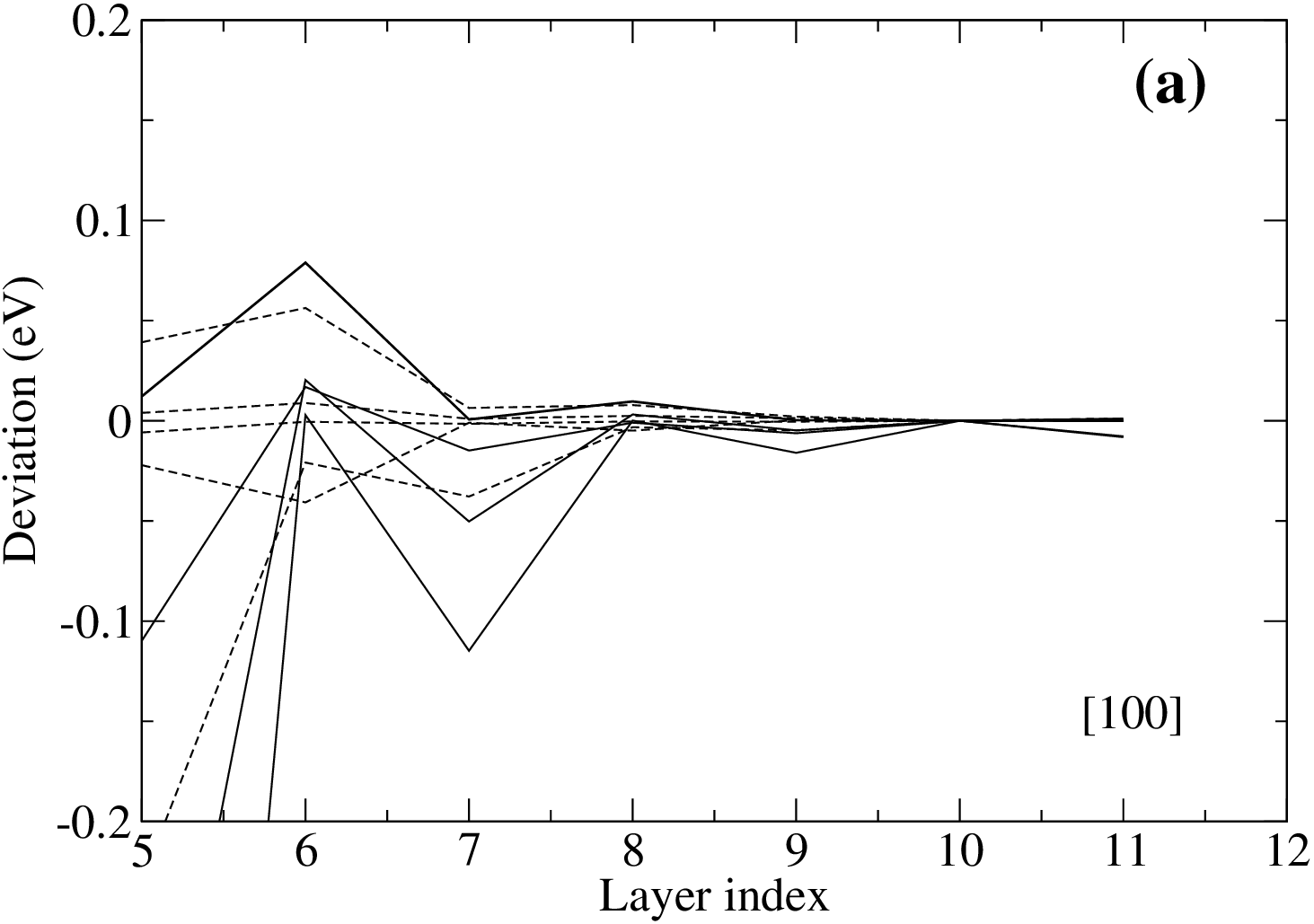} &
\includegraphics[width=2.8in,clip]{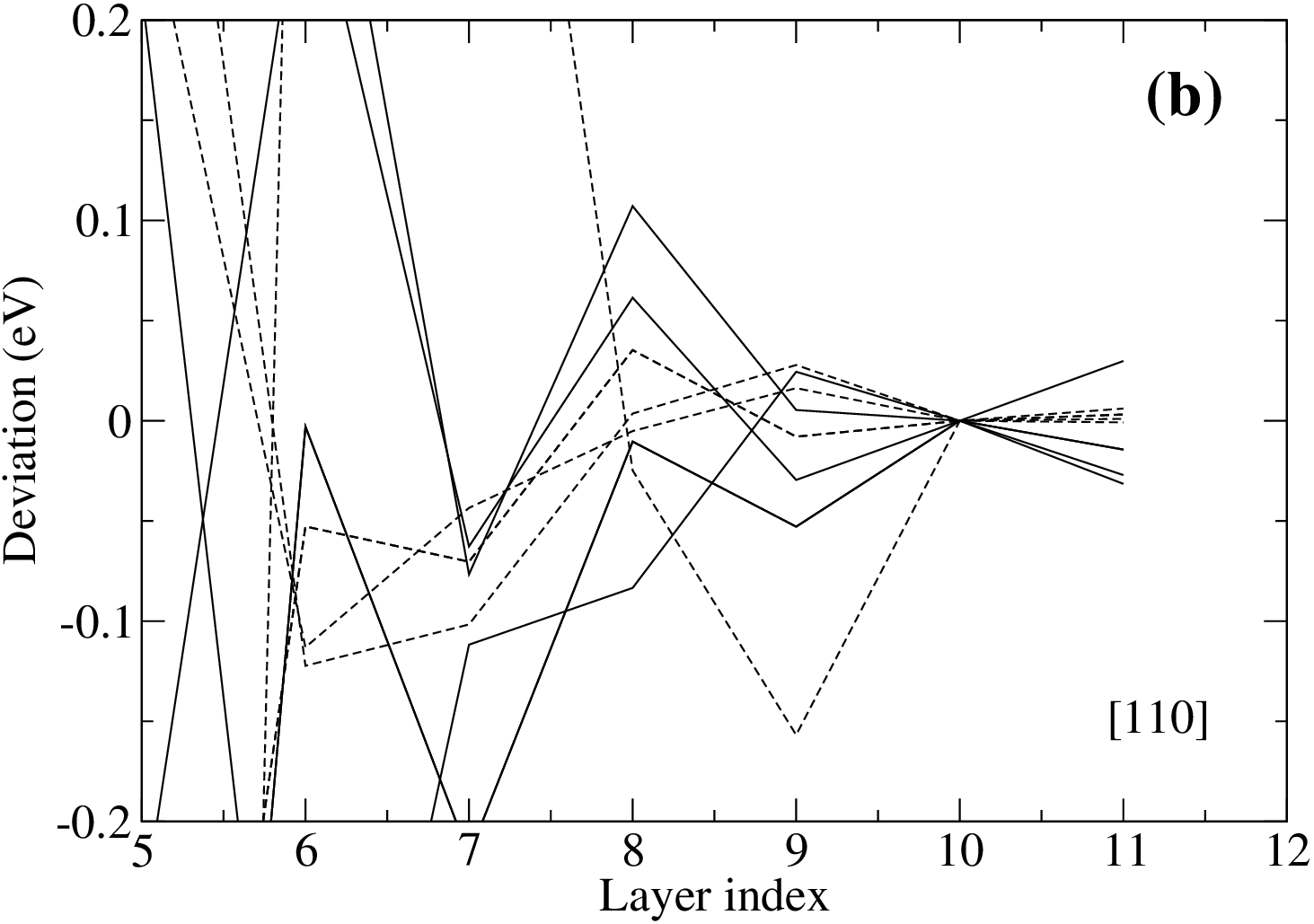} \\
\includegraphics[width=2.8in,clip]{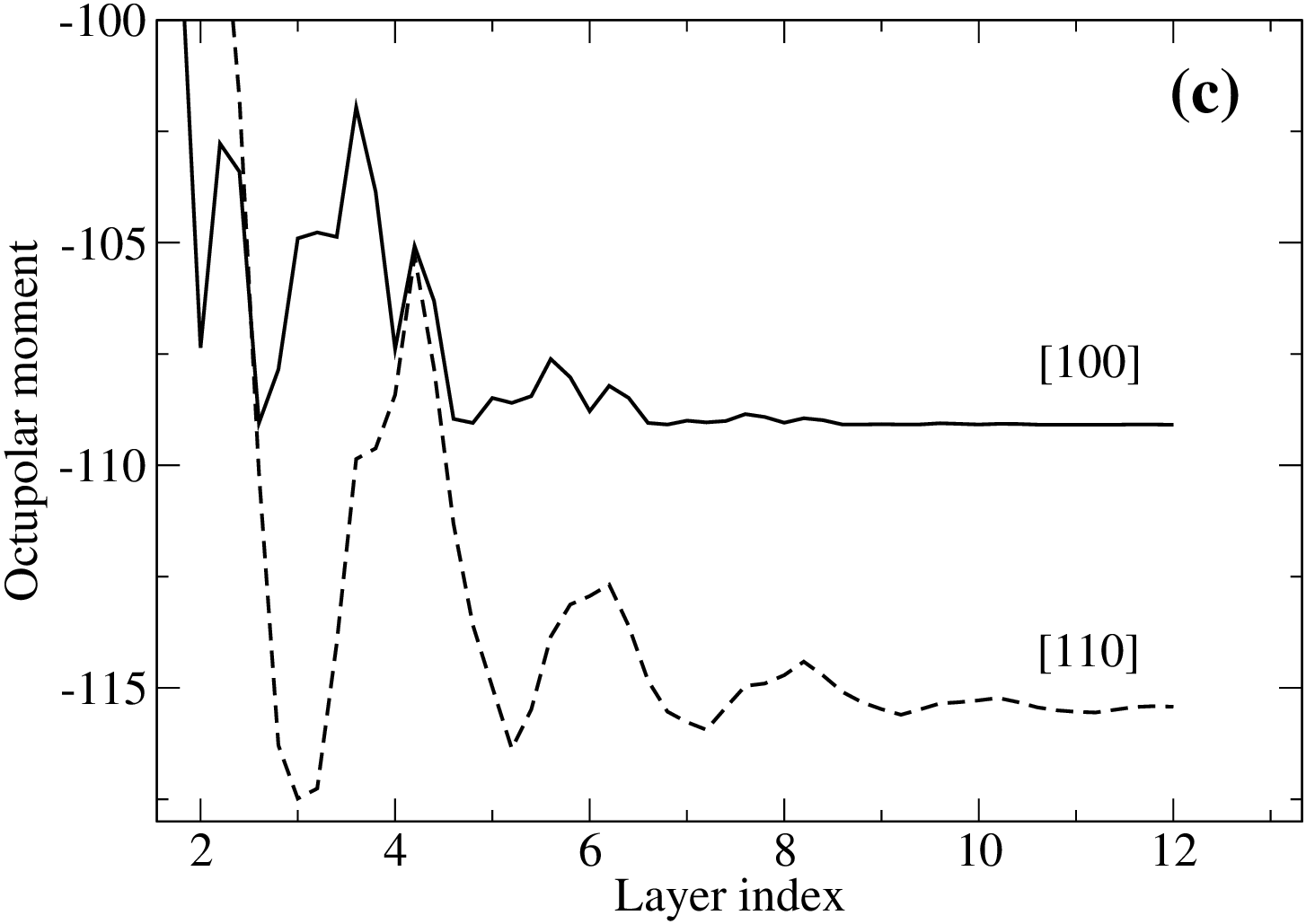} &
\includegraphics[width=2.8in,clip]{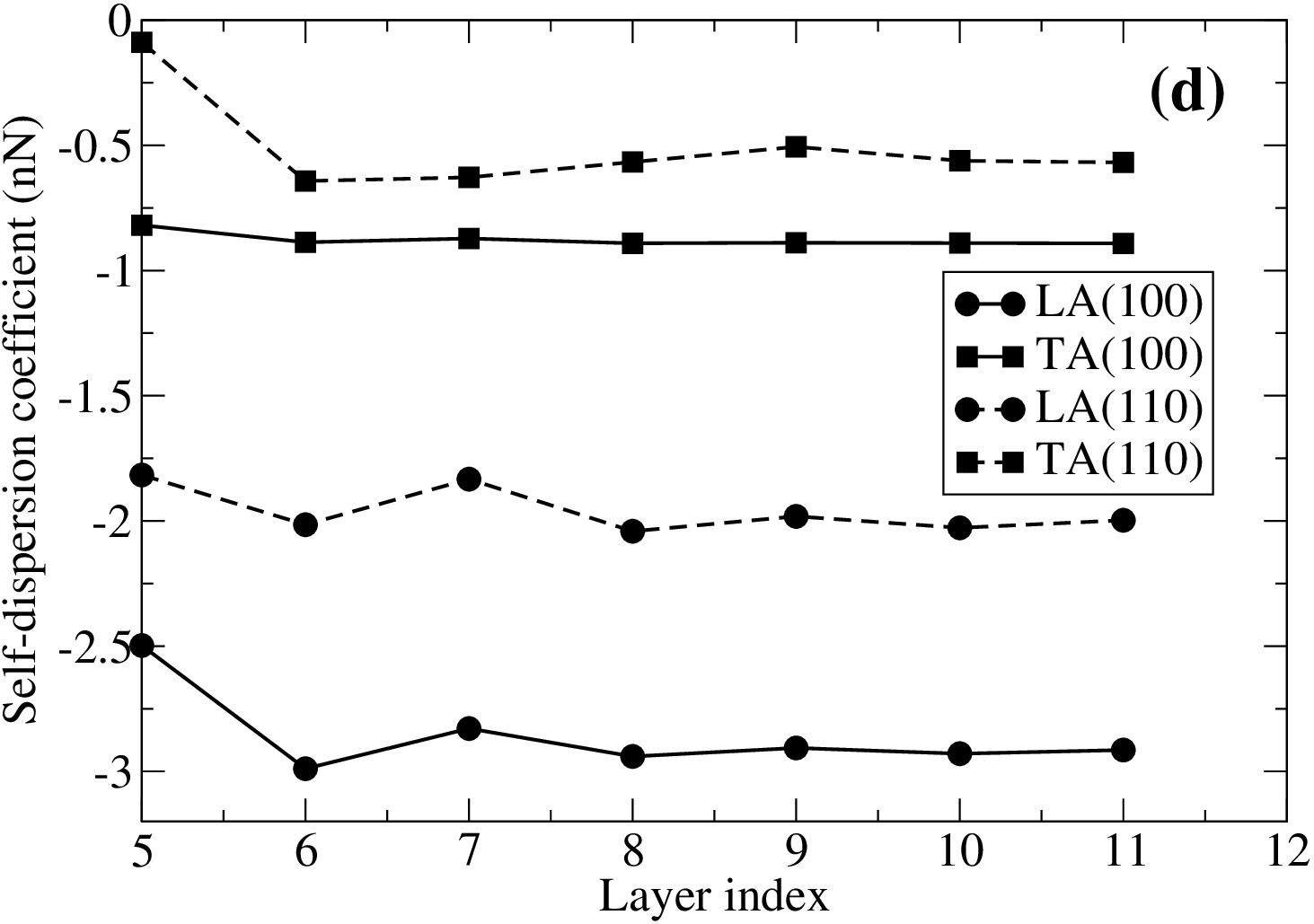} 
\end{tabular}
\end{center}
\caption{ Convergence of various quantities with the real-space cutoff of the
interatomic force constants along a given direction. (a-b): Flexocoupling coefficients along [100] and [110];
solid and dashed lines refer to longitudinal and transverse modes, respectively; the reported values are
the deviations with respect to the $n=10$ point. (c): Electronic octupolar moments; the converged
values are $\tilde O_{100}=-109.1$ a.u. and $\tilde O_{110}=-115.4$ a.u. (to be compared with 
$\tilde O_{100}=-108.8$ a.u. and $\tilde O_{110}=-115.3$ a.u., obtained in Ref.~\onlinecite{artcalc}). 
(d) Electronic contribution to the fourth-order dispersion ($\tilde h^{\rm el}$). \label{fig1} }
\end{figure*}

Our calculations are performed within the local-density approximation~\cite{Perdew/Wang:1992} 
to density-functional theory. The interactions between valence electrons and ionic cores are described by 
separable norm-conserving pseudopotentials in the Troullier-Martins~\cite{troullier} 
form, generated with the fhi98PP code.~\cite{fhi98pp}
The reference states (the numbers in brackets
indicate the core radius in bohr) of the isolated neutral atom used for the
generation of the pseudopotentials are $2s$(1.4), $2p$(1.4) and $3d$(1.4) for O, 
$4s$(1.5), $4p$(1.5) and $4d$(2.0) for Sr and $3s$(1.3), $3p$(1.3) and $3d$(1.3) for Ti. The local
angular-momentum channel is $l = 2$ for Sr and O, $l = 0$ for Ti. 
The cutoff for the wavefunction plane-wave basis is
set to 300 Ry to ensure optimal accuracy in the numerical differentiations in $q$-space.
The surface Brillouin zone of the SrTiO$_3$ primitive cell is sampled by means of 
a $12 \times 12 \times 12$ Monkhorst-Pack mesh. The long-wave expansion of the
dynamical matrix is performed via the following procedure.

First, we calculate the full dynamical matrix, by means of density-functional
perturbation theory~\cite{Gonze,Gonze/Lee,Baroni/deGironcoli/DalCorso:2001} 
as implemented in ABINIT,~\cite{abinit} on a regularly
spaced stripe of ${\bf q}$ points in reciprocal space.
Compatibly with the chosen $k$-point set, we use $\Gamma$-centered stripes of 12 
points spanning a line in reciprocal space, either along [100] or [110]. 
(The 
dynamical matrix at $\Gamma$ is corrected with the nonanalytical term that corresponds
to the direction in ${\bf q}$-space under study, which we separately calculate by
means of a standard electric field response calculation.)
Second, we operate a one-dimensional Fourier transform on each matrix element,
which provides us with the real-space force constants along a given direction.
Such force constants decay \emph{exponentially} in real space, and their
moments can be therefore calculated very accurately. The convergence of
any quantity with respect to the real-space cut off of the interatomic constants
can be also easily monitored.
These moments provide us with the desired long-wave expansion terms of
the $\hat{D}$-matrices (i.e. those with the nonanalytic electrostatic 
terms included).
Next, parallel with the analysis of the interatomic force constants we perform an
analogous Fourier processing of the induced charge density, which provides us
with the electronic octupolar moments, and hence with the longitudinal
components of the flexoelectric tensor. 
Finally, by using the known relationships between short-circuit and open-circuit
flexoelectric response, we appropriately combine the charge octupoles 
and the calculated $\hat{D}$-matrices to extract the full flexocoupling
tensor components and, in turn, all the necessary quantities to study SGE
and flexoelectricity in bulk SrTiO$_3$.

In Table~\ref{tab1} we report the calculated values of a few standard
lattice-dynamical and dielectric properties of bulk SrTiO$_3$: the optical mode 
frequencies, their associated dynamical charges and the dielectric
constant (both in the static and high-frequency limits). 
These quantities are shown here both for reference, and also because they are
directly involved in the higher-order tensors describing the strain-gradient 
response of the crystal.
To calculate the latter, and thereby demonstrate the formalism developed in this
work, a number of additional basic ingredients are needed: the flexocoupling
coefficients ($\tilde f$), the electronic octupolar moments, and the relevant 
frozen-ion SGE coefficients ($\tilde h$).
%
Since these quantities are calculated as a real-space moment of some
Fourier-transformed lattice-dynamical quantity, one must choose a
cutoff distance beyond which the lattice sum (or the integral) is 
truncated.
The convergence of each of the aforementioned quantities with
respect to such a cutoff (expressed in number of atomic monolayers) 
is shown in Fig.~\ref{fig1}. In all cases the convergence is excellent,
e.g., it is of the order of 0.1 eV (i.e., well below 1\%) in the flexocoupling 
coefficients along [110], and even (much) better in the [100] case. 
We shall initially report the values of the aforementioned quantities as
calculated under ``mixed electrical
boundary conditions'' (MEBC).~\cite{Hong-11} (longitudinal modes 
experiences an open-circuit environment, while short-circuit is naturally 
imposed by the periodicity of the lattice in the transverse plane),
and later discuss how to recast them in a tensorial form by separating
the electrostatic contribution.
Consequently, the octupolar moments, $\tilde O_{\hat{\bf q}}$, reported in 
Fig.~\ref{fig1} are related to the longitudinal component of the frozen-ion 
flexoelectric tensor by $\mu_{\hat{\bf q}} = \epsilon_\infty \tilde O_{\hat{\bf q}} / 6 \Omega$.

%
%

\subsection{Flexocoupling coefficients in MEBC}

\begin{table}
\begin{center}
\begin{ruledtabular}
\begin{tabular}{c @{\hspace{15pt}} rr @{\hspace{15pt}}  rr }
    & \multicolumn{2}{c}{$[100]$}  & \multicolumn{2}{c}{$[110]$}    \\
                 &   \multicolumn{1}{c}{L}       &   \multicolumn{1}{c}{T}     & 
                     \multicolumn{1}{c}{L}       &   \multicolumn{1}{c}{T}                        \\
\hline 
 A               &    137.13  &    43.46  &    132.02  &         48.59             \\
 1               &  $-$83.45  & $-$44.53  &  $-$65.53  &      $-$27.89             \\
 2               &    108.45  &     5.84  &     52.15  &         29.92             \\
 3               & $-$155.16  & $-$22.87  & $-$100.36  &      $-$89.90             \\               
 S               &      0.00  &    43.70  &     57.33  &      $-$13.65           
\end{tabular}
\end{ruledtabular}
\end{center}
\caption{ Calculated energy flexocoupling coefficients in MEBC, corresponding to
acoustic phonon modes propagating along [100] and [110].
Labels refer to the self-coupling of the acoustic mode (A), to the
IR-active optical modes (1-3) and to the silent mode (S). 
L and T indicate longitudinal and transverse polarization, respectively. 
The character of the IR-active modes is consistent with the L or T label.
Values are in eV units. 
\label{tab2}}
\end{table}

The central quantity that one needs when dealing with either
flexoelectricity or strain-gradient elasticity is the flexocoupling
tensor -- for this reason we shall describe its calculation in 
detail. 
The first step, which will be outlined in this section, 
is the calculation of the longitudinal and transverse flexocoupling 
coefficients, $\tilde{f}^{\rm L,T}_{\hat{\bf q}}$, along the [100] or [110] 
direction in ${\bf q}$-space.
These are given by the second moments (along the direction $\hat{\bf q}$) 
of the ``bare'' dynamical matrix, $\hat{D}$, i.e., with the 
electrostatic interactions included; this means that MEBC are
naturally imposed along $\hat{\bf q}$.

One must keep in mind that the coefficients that
one obtains this way are specialized to the direction 
$\hat{\bf q}$ and to the polarization (longitudinal or transverse) 
of the mode: For example, some of the $\tilde f$ coefficients
describe the interaction between longitudinal acoustic (LA) and 
longitudinal optic (LO) modes ($\tilde f^{\rm L}$), while others couple 
transverse acoustic (TA) modes to transverse optic 
(TO) phonons ($\tilde f^{\rm T}$).~\footnote{
  Since we are dealing with high-symmetry directions, 
  the two subspaces of the longitudinal and transverse phonons
  are decoupled, and can be treated independently.}
As TO and LO modes experience dissimilar electrical boundary 
conditions, they differ \emph{even at the Brillouin zone center}; 
this implies that $\tilde f^{\rm L}$ coefficients cannot be
mixed or compared to $\tilde f^{\rm T}$ coefficients, let
alone treated as the components of a single tensor. (The practical 
procedure to extract a proper tensorial expression will be discussed 
shortly.)

The calculated values of the $\tilde{f}_{\hat{\bf q}}$ coefficients
are reported in Table~\ref{tab2}.
In addition to the coupling to the IR-active modes, 
which are sensitive to the above considerations on the electrical 
boundary conditions, we also show the ``self-coupling'' 
of the acoustic branch (these directly relate to the relevant 
component of the elastic tensor), and the coupling
to the ``silent'' (S) mode. 
The latter, of course, does not carry a dynamical dipole and is therefore
irrelevant for flexoelectricity; still, as we shall see in the following
Section, it does contribute to strain-gradient elasticity.

\subsection{Acoustic phonon dispersion}

\label{sec:disp}

\begin{table}
\begin{center}
\begin{ruledtabular}
\begin{tabular}{c @{\hspace{15pt}} rr @{\hspace{15pt}}  rr }
    & \multicolumn{2}{c}{$[100]$}  & \multicolumn{2}{c}{$[110]$}    \\
                 &   \multicolumn{1}{c}{L}       &   \multicolumn{1}{c}{T}     & 
                     \multicolumn{1}{c}{L}       &   \multicolumn{1}{c}{T}                        \\
\hline 
 A   &  $-$2.93 &     $-$0.89  &    $-$2.03   &        $-$0.56     \\    
1    & $-$10.76 &    $-$62.69     & $-$6.64       &   $-$24.59     \\
2    & $-$2.35  &     $-$0.05    &  $-$0.54       &    $-$1.29    \\
3    & $-$1.61  &     $-$0.07    &  $-$0.67       &    $-$1.09    \\
S    &    0.00  &     $-$1.46    &  $-$2.50       &    $-$0.14   \\
\hline
Total   & $-$17.65 &     $-$65.16    & $-$12.39 &    $-$27.67  
\end{tabular}
\end{ruledtabular}
\end{center}
\caption{  Contributions to the dispersion of the acoustic 
branches in a vicinity of $\Gamma$, corresponding to the $\tilde h$ coefficient
defined in the text.
Longitudinal and transverse phonon modes propagating along [100] and [110] are considered.
Labels correspond th the self-dispersion (A), IR-active optical modes (1-3) and silent
mode (S). Values are in nN.
\label{tab3}}
\end{table}

In Table~\ref{tab3} we report the calculated values of the $\tilde{h}$
coefficients, referring to nonlocal elastic effects in MEBC.
These coefficients are further decomposed into a purely electronic 
(self-dispersion) term, which we shall indicate as ``frozen-ion'' (FI)
hereafter, and a number of lattice-mediated (LM) contributions,
which are associated to the relaxation of each zone-center 
optical modes, either IR-active or silent.
(Such a decomposition is in all respects equivalent to the
better-known case of linear elasticity, where the corresponding
materials constants are also conveniently split into a FI
and a LM contribution.)
%
It is clear from the table that all the values are negative,
i.e. both effects lead to a systematic \emph{softening} of the 
elastic response of the crystal at short length scales. 
The physical mechanisms that lie behind this observation are 
quite dissimilar in the FI and LM cases, so I shall discuss them 
separately in the following, starting from the former.

\begin{figure}[!b]
\begin{center}
\includegraphics[width=2.8in,clip]{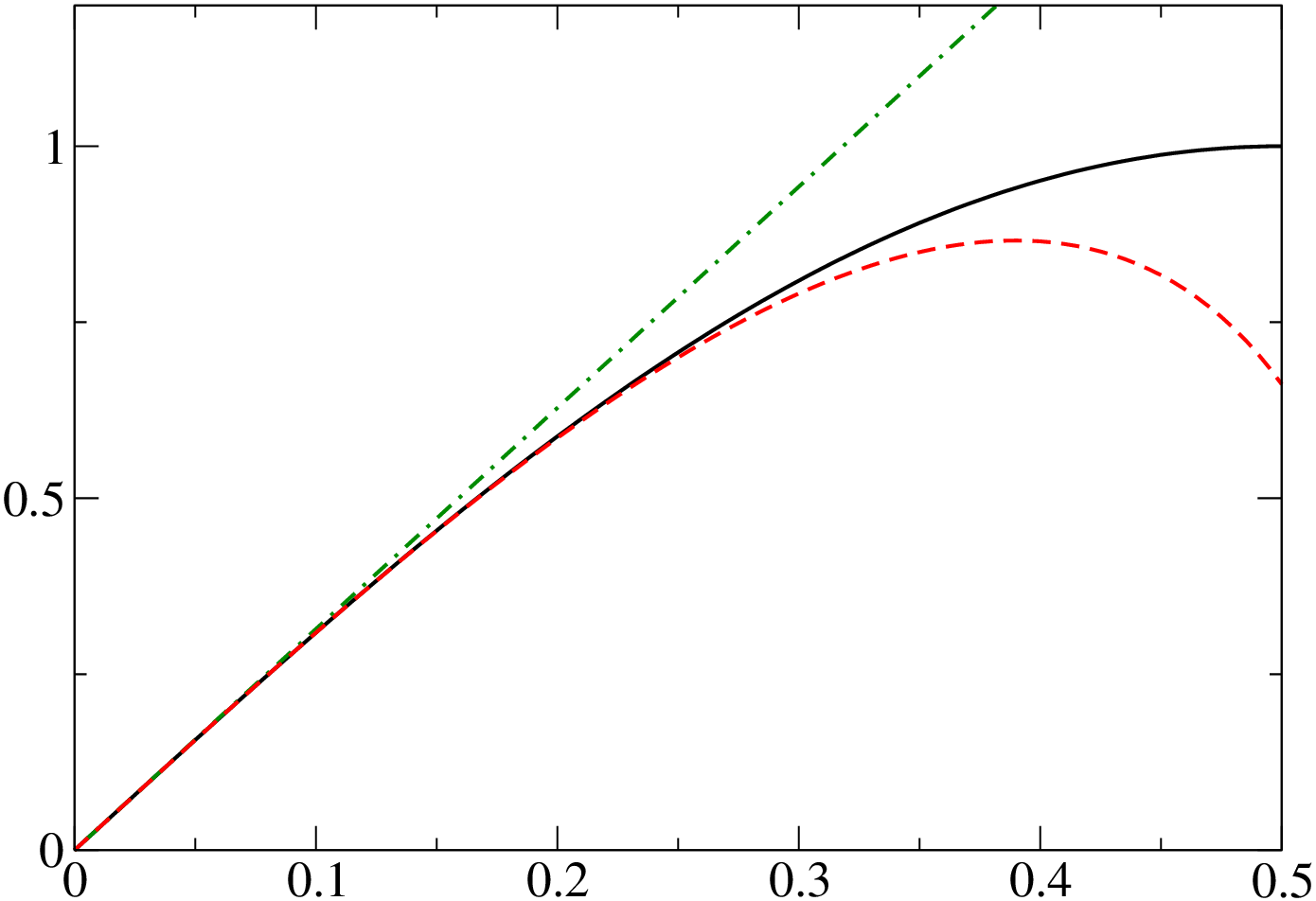}
\end{center}
\caption{\label{figchain} Dispersion relation of a linear chain
(black curve) as approximated by local elasticity (green dot-dashed)
and SGE (red dashed).}
\end{figure}


\begin{table}
\begin{center}
\begin{ruledtabular}
\begin{tabular}{c rrrr}
 &          \multicolumn{2}{c}{First-principles} &  \multicolumn{2}{c}{Model} \\   
 & \multicolumn{1}{c}{L} & \multicolumn{1}{c}{T} &\multicolumn{1}{c}{L} & \multicolumn{1}{c}{T} \\
\hline 
 $[100]$    &  $-$2.93   &   $-$0.89  &   $-$4.77 & $-$1.51    \\    
 $[110]$    &  $-$2.03   &   $-$0.56  &   $-$2.30 & $-$0.85    \\
 $[111]$    &  $-$1.87   &   $-$0.38  &   $-$1.51 & $-$0.54  
\end{tabular}
\end{ruledtabular}
\end{center}
\caption{ Frozen-ion contribution to the SGE coefficients in MEBC
(``first-principles'') compared with a rough estimation based on the
1st-neighbor lattice model described in the text (``model'').
The model values were obtained by setting the $a$ parameter to the
periodicity of the lattice along the phonon directions, i.e. 
$a_0$, $a_0 / \sqrt{2}$, and $a_0 / \sqrt{3}$ respectively along [100],
[110] and [111]. Values are in nN.
\label{tab3a}}
\end{table}

To understand the origin of the self-dispersion of the acoustic 
branches it is instructive to consider the simple textbook
model of a linear chain of atoms interacting with first-neighbor 
springs.
The dispersion of the LA branch is trivially given
by
\begin{equation}
\omega^2(q) = 2 \frac{k}{m} [1 - \cos(qa)],
\end{equation}
where $k$ is the spring constant, $m$ is the mass and
$a$ is the lattice spacing. 
By performing a long-wave expansion to $\mathcal{O}(q^4)$,
analogously to the procedure used in the remainder of this
work, one readily obtains a continuum energy functional for
this system,
\begin{equation}
E = \frac{q^2}{2} \mathcal{C} + \frac{q^4}{2} \tilde h,
\end{equation}
where the elastic and hyperelastic constants are
\begin{eqnarray}
\mathcal{C} &=& \, \, \, \frac{k a^2}{\Omega}, \\
\tilde h    &=& -\frac{k a^4}{12\Omega}.
\end{eqnarray}
(We have introduced the volume factor $\Omega$ by supposing that
the chain of atoms is, in fact, a chain of atomic \emph{planes},
consistent with the three-dimensional nature of the SrTiO$_3$
crystal under study.)
In Fig.~\ref{figchain} we show a comparison of the 
phonon dispersion as predicted by the continuum SGE functional
with the exact discrete reference.
This analysis allows us to relate the two elastic coefficient as
\begin{equation}
\tilde h = -a^2 \mathcal{C} / 12.
\label{eqchain}
\end{equation} 
This result implies that
$\tilde h$ is primarily due to the discreteness of the 
lattice, and will produce measurable effects at a lengthscale
that is comparable to the interatomic spacing, $a$.
While SrTiO$_3$ is undoubtedly more complicated than this 
toy model, it is interesting to compare the predictions of Eq.~(\ref{eqchain})
with the actual values of $\tilde h$ calculated from first-principles,
to see if, at least qualitatively, the above ideas are correct.
As one can readily appreciate from Table~\ref{tab3a}, the two
sets of values display a consistent trend, and even quantitatively
they lie within a factor of two in all cases, confirming that we
are indeed on the right track.
Such an agreement tells us that the FI contribution to strain-gradient 
elasticity is utterly small, and becomes relevant only at a lengthscale
that is comparable to the interatomic spacing. (Similar conclusions
were drawn in Ref.~\onlinecite{Maranganti-07}.)
Its inclusion in a continuum thermodynamic functional appears therefore 
of limited interest, except for guaranteeing the gauge invariance of
the theory as we shall see in Section~\ref{sec:gi}.

The LM contribution, related to the optical modes, is negative by
construction, and in the transverse cases is largely dominated
by the ferroelectric ``soft'' mode. (In the longitudinal
case, the overall value of $\tilde h$ is more equally distributed.)
%
%
%
That the soft mode plays a dominant role in $\tilde h$ is
no surprise, given its very low transverse frequency (recall that 
the squared frequency appears at the denominator in the SGE energy) in 
our computational model of SrTiO$_3$.
After the inclusion of the LM contributions, the resulting characteristic length scales 
(usually defined in the literature as $\xi = \sqrt{\mathcal{C}/|\tilde h|}$), are significantly larger compared to the 
previous estimation of $\xi \sim a / \sqrt{12} $, obtained at the frozen-ion level.
Still, the value of $\xi$ hardly reaches 1 nm in the 
present first-principles model of SrTiO$_3$, questoning again 
the general relevance of the SGE (and flexoelectric) energy
in continuum simulations of macroscopic phenomena.
It is important, however, to emphasize a notable consequence of
the theory presented so far: the above length scale \emph{diverges}
near a ferroelectric phase transition, i.e. when the 
frequency of the soft mode tends to zero. This suggests that
SGE may lead to interesting physical effects whenever an
optical phonon undergoes a critical behavior, and that 
lattice-mediated flexoelectric/SGE effects cannot \emph{a priori} be 
neglected in such a regime.


%
%
%
%
%
%

\subsection{Macroscopic coupling tensors}

\begin{table}
\begin{center}
\begin{ruledtabular}
\begin{tabular}{ c  r  r  r  }
      &  $C_{xx,xx}$ & $C_{xy,xy}$ & $C_{xx,yy}$ \\
\hline
 Dynamical matrix &   386.2  &   122.4  &   112.6 \\ 
 Strain           &   386.2  &   122.4  &   112.6  
\end{tabular}
\end{ruledtabular}
\end{center}
\caption{Calculated elastic tensor of bulk SrTiO$_3$. Values in the upper row were obtained by
using the dynamical matrix approach described in this work. The lower row was obtained by taking
finite differences of the calculated stress tensor while varying the strain around the equilibrium cubic 
configuration. Values are in GPa. \label{tab4}}
\end{table}

In this Section we shall proceed to extracting, from the results 
presented so far, the elastic and flexocoupling coefficients in a
proper tensorial form. 
Regarding the elastic tensor, it can be trivially extracted 
from the [100] and [110] ``flexocoupling'' coefficients of the 
acoustic mode with itself. 
As there are four calculated values and three independent entries, the
redundancy can be used as a consistency check. A second numerical
test consists in comparing the values calculated this way to a
more standard calculation of $\bm{\mathcal{C}}$, performed via
finite differences in the strain.
As one can see from the results reported in Table~\ref{tab4},
the two procedures show essentially perfect agreement: deviations
are smaller than 0.1 GPa in all cases.

Recasting the $\tilde f$ coefficient into a tensorial form is more 
delicate, and requires two preliminary steps: (i) the nonanalytic electrostatic 
terms need to be removed from $\hat{D}^{(2)}$, thereby obtaining $\mathcal{D}^{(2)}$;
(ii) the basis of zone center-eigenmodes on which $\mathcal{D}^{(2)}$ is projected
need to be calculated under isotropic short-circuit conditions, rather than MEBC.
Then, just like in the elastic case, we have three independent entries and
four independent values for each optical mode; this is again a stringent
test of the overall consistency of the implementation.
(In practice, we treat the [100] values as exact, and average the error on the
[110]-related terms. The deviation is very small, of the order of 0.1--0.2 eV.)
The resulting values, which are one of the main results of this work, 
are reported in Table~\ref{tab5}.

Note, first of all, the strong reference dependence of the individual 
coefficients, which can even change sign in some cases when going from
a $p$-type to a $n$-type regime.
(We use ``$p$-type'' and ``$n$-type'' as shortcuts to indicate that 
either the valence-band edge or the conduction-band edge was chosen as 
the reference potential.)
What this really means physically is that, if we think of SrTiO$_3$ 
as a doped semiconductor, the coupling between strain gradients and
zone-center optical phonons will strongly depend on the character of
the majority carriers (electrons or holes). 
If SrTiO$_3$ is in a perfectly insulating state, on the other hand,
the choice of one or the other reference is completely arbitrary -- 
what changes is just the physical meaning of the ``electrostatic
potential'' that stems from a self-consistent solution of the
electromechanical problem.

\begin{table}
\begin{center}
\begin{ruledtabular}
\begin{tabular}{c @{\hspace{15pt}} rr @{\hspace{15pt}}  rr @{\hspace{15pt}} r}
    & \multicolumn{2}{c}{$f_{xx,xx}$}  & \multicolumn{2}{c}{$f_{xx,yy}$} & \multicolumn{1}{c}{$f_{xy,xy}$}   \\
    &  $n$-type & $p$-type              &  $n$-type  &  $p$-type            &                                     \\
\hline 
 1   &  $-$51.1  &  $-$90.2  &                  5.1  &  $-$34.0             &   $-$44.5  \\
 2   &     74.4  &     64.1  &                 14.8  &      4.5             &       5.8  \\
 3   & $-$181.6  & $-$201.7  &               $-$1.4  &  $-$21.6             &   $-$22.9  \\                 
 S   &      0.0  &      0.0  &                 27.3  &     27.3             &      43.7  
\end{tabular}
\end{ruledtabular}
\end{center}
\caption{ Calculated type-II energy flexocoupling coefficients (in eV units). For the longitudinal
$(xx,xx)$ and transverse $(xx,yy)$ components, both the $n$-type and $p$-type values are shown,
while the shear $(xy,xy)$ coefficient is reference-independent. The three components are
often indicated in the literature as $f_{11}$, $f_{12}$ and $f_{44}$, respectively.
\label{tab5}}
\end{table}

Not all the coupling coefficients are affected by such a reference dependence, though:
The shear components $f^j_{xy,xy}$ (also known in the literature
as $f_{44}$) are unsensitive to this arbitrariness. A closer look allows 
us to identify an additional linear combination of the $f$-coefficients 
where the ambiguity cancels out, 
\begin{equation}
f^{\rm T}_{110} = \frac{1}{2} ( f_{11} - f_{12} ),
\end{equation}
which is relevant for a transversally polarized (i.e., with the displacement 
vector oriented along [$1\bar 1 0$]) acoustic phonon propagating along [110].
Transverse phonons along any conceivable direction are described
by a linear combination of the $f_{44}$ and $f^{\rm T}_{110}$ coefficients, 
and the reference independence is consistent with the preservation
of translational periodicity along the displacement direction.
Regarding the actual values, in the case of the soft mode (TO1) we obtain
\begin{eqnarray}
\frac{f^{\rm TO1}_{44}}{Z^*_{\rm TO1}} \, \,  &=&  -1.96 \, \, {\rm V} \label{f44} \\
\frac{f^{\rm TO1}_{11} - f^{\rm TO1}_{12}}{2Z^*_{\rm TO1}} &=& -1.24 \, \, {\rm V}.
\label{f11f12}
\end{eqnarray}
(We converted the flexocoupling coefficients to voltage units by 
dividing them by the mode dynamical charge for a better comparison
with existing literature data.)
These values seem to be in 
overall agreement with the existing experimental 
estimates ($|f_{11}-f_{12}| =$1.2--1.4 V, $|f_{44}|=$1.2--2.4 V)~\cite{Kvasov-15,Yudin-13,pavlo_review,Brillouin}.

An independent first-principles calculation of such quantities was 
recently reported in Ref.~\onlinecite{Kvasov-15}. Our results present
significant quantitative differences, especially regarding the
[110] coefficient (a value of -0.2 V was reported by Kvasov and Tagantsev).
Such a discrepancy may be in part due to differences in the general 
computational setup (e.g. exchange and correlation functionals, 
pseudopotentials), but also in the specific procedure that one
uses to extract the $f$-tensor from the linear-response data.
We stress that a correct treatment of the electrical boundary
conditions, as we have extensively discussed in the course of 
this work, is essential for a reliable calculation of ${\bf f}$.
%
%
Interestingly, if we were to estimate the transverse components
of ${\bf f}$ from the TA dispersion curves (by assuming, 
following Ref.~\onlinecite{Yudin-13}, that TO1 is the dominant 
source of curvature of the branch), 
we would make an error of 2\% and 6\%, respectively in
the [100] and [110] coefficient (this can be easily inferred
from the data of Table~\ref{tab3}). 

\subsection{Gauge invariance of LA phonons}

\label{sec:gi}

\begin{table}
\begin{center}
\begin{ruledtabular}
\begin{tabular}{c rrr }
                 &   \multicolumn{1}{c}{$n$-type}       &   \multicolumn{1}{c}{$p$-type}     & 
                     \multicolumn{1}{c}{$\varphi$}                          \\
\hline 
%
A   &    $-$3.043      &  $-$2.934    &    $-$17.276      \\    
1    & $-$257.089  &     $-$82.421    &  $-$5693.678     \\
2    &   $-$5.926    &    $-$7.985    &     $-$0.828    \\
3    &   $-$5.488    &    $-$4.449    &    $-$18.815     \\
El.  &    253.899    &      80.142    &     5712.950 \\
\hline
Total   & $-$17.647 &    $-$17.647    &    $-$17.647

\end{tabular}
\end{ruledtabular}
\end{center}
\caption{  Decomposition of the dispersion of the LA phonon along [100]
into self-dispersion (A), optical modes (1-3) and electrostatic 
(El.) contributions. Three different assumptions for the short-circuit
boundary conditions are shown: $p$-type screening (flat valence band),
$n$-type screening (flat conduction band) and electrostatic screening
(flat macroscopic electrostatic potential, $\varphi$).
The overall result is independent of this
choice, and coincides with the value ($-$17.646 nN) calculated under
open-circuit conditions (see Table~\ref{tab3}). Values are in nN.
\label{tab6}}
\end{table}

It is useful, before closing this long Section, to perform a 
further consistency check of the formalism, this time by 
focusing on the gauge invariance. 
Apart from the obvious validation purposes, this exercise will
provide a quantitative flavor on exactly \emph{how much} the reference potential
ambiguity affects the partition between SGE and Maxwell energy.
As a representative example, I will focus on the dispersive
behavior of the LA phonon branch along [100], whose analysis 
has already been presented in the first column of Table~\ref{tab3}.
In Table~\ref{tab3}, however, the total $\tilde h$ coefficient 
was decomposed into the contributions from the LO modes
and the open-circuit self-dispersion of the LA branch. 
Here I shall, instead, decompose the same value into 
contributions from TO modes, the \emph{short-circuit} self-dispersion
of the branch [as given by Eq.~(\ref{hbarsc})], and the Maxwell energy of the 
flexoelectrically induced electric fields.
Of course, depending on the choice of the reference potential, the
individual pieces will vary but the overall sum must remain the same.

The results of this new decomposition, performed for three different 
choices of the reference potential are shown in Table~\ref{tab6}. 
(Next to the $p$-type and $n$-type results,  
I also show a decomposition performed by using the bare electrostatic 
potential as a reference -- the corresponding column is marked as $\varphi$.)
The contribution of the optical phonons, as expected, is largely dominated
by the soft mode (TO1). 
Such a contribution, which is negative definite, strongly depends on
the reference, and becomes very large in the case of the bare electrostatic
reference.
This negative term, however, is almost
exactly cancelled in all cases by an equally large and positive contribution
from the Maxwell energy.
The overall sum, which depends on the slight discrepancy between these two values
and on the (much smaller) residual contribution from self-dispersion and 
other optical modes, is gauge-independent as expected, and accurately 
matches the value reported in Table~\ref{tab3}.

This analysis highlights two important facts that were already
anticipated earlier. First, a consistent description of strain-gradient
elasticity is necessary for building a well-defined functional that
incorporates flexoelectric effects.
Second, insisting on choosing the electrostatic potential as a 
reference, as implicitly assumed in earlier \emph{ab initio}
works,~\cite{artcalc,Hong-13} may lead to an awkward partition of 
the energetics between two extremely large terms, which are opposite 
in sign and almost exactly cancel.
(This is unpalatable in practical implementations of the theory,
as numerical errors might affect the two terms in a dissimilar
way, and thus be artificially amplified.)
This corroborates the arguments of Ref.~\onlinecite{adp}, 
where the choice of the valence and conduction band edges 
as a reference when modeling flexoelectric phenomena was advocated
for closely related reasons.
Since calculations of flexoelectricity are usually performed
(as in this work) in the framework of density-functional theory,
adopting the valence-band edge as the energy reference appears as
the most sensible choice: This is the only band energy that
is, in principle, correctly described within ``exact'' DFT,
while the physical meaning of other single-particle eigenvalues 
(including the conduction-band minimum) is less clear.

\begin{figure*}
\begin{center}
\includegraphics[width=4.5in,clip]{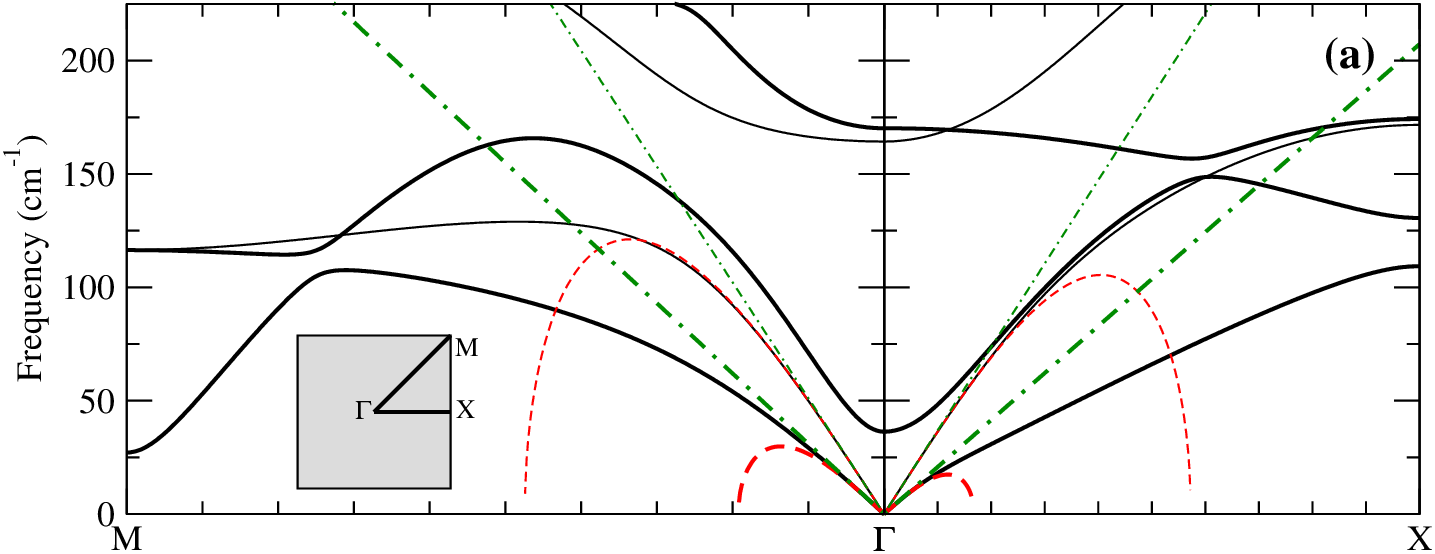} \\
\vspace{10pt}
\includegraphics[width=4.5in,clip]{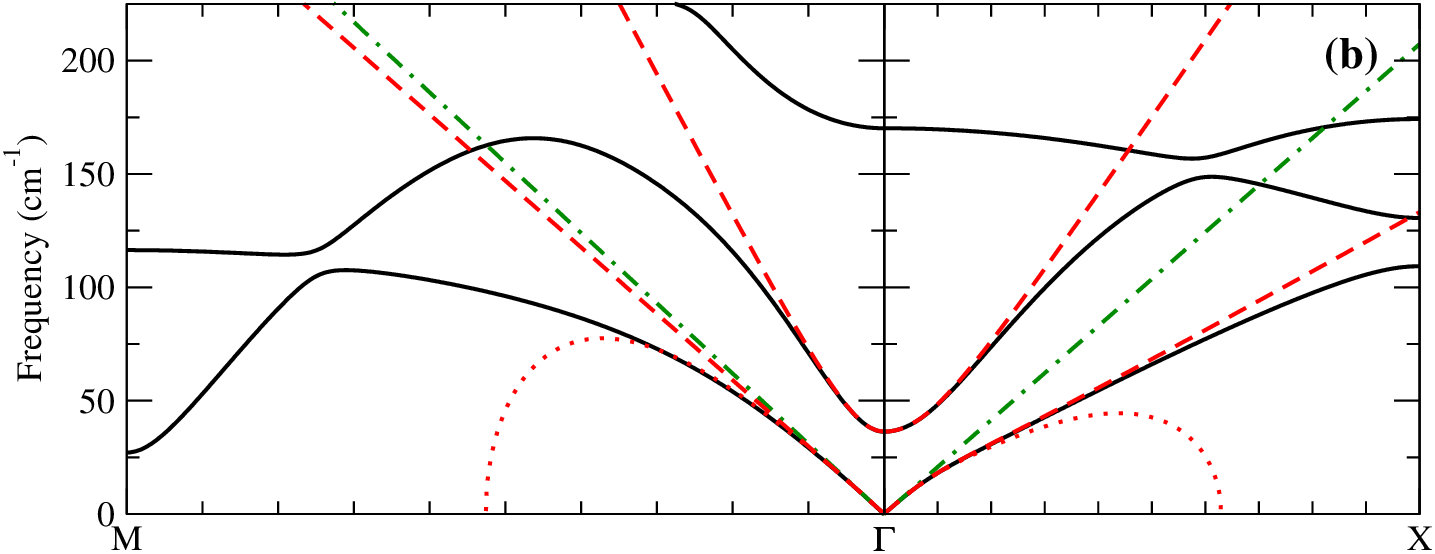} \\
\end{center}
\caption{ (a): Lattice-dynamical analysis of the continuum strain-gradient theory
of Section~\ref{sec:continuum}, applied to the SrTiO$_3$ case.
Black solid curves represent the dispersion of the
Fourier-interpolated \emph{ab initio} phonon frequencies (thin and thick correspond, respectively,
to longitudinal and transverse branches); green dot-dashed lines correspond to the linear
dispersion given by classical elasticity; red dashed curves correspond to 
the continuum model. 
The inset indicates the paths in the 2D Brillouin zone that correspond to the 
reported phonon branches. (b): Revised continuum functional of Eq.~(\ref{newfunc}),
either with (dotted curves) or without (dashed curves) the ``background'' SGE 
term; only transverse modes are shown.
\label{fig2} }
\end{figure*}


%


\section{Discussion}

\label{sec:discuss}


\begin{table}
\begin{center}
\begin{ruledtabular}
\begin{tabular}{c rrr }
        &   $\Omega \mathcal{C}$ & $f$  & $g$ \\ 
\hline 
$[100]$ &  43.46 &  $-$44.53  &   94.26  \\
$[110]$ &  48.59 &  $-$27.89  &  199.57 
\end{tabular}
\end{ruledtabular}
\end{center}
\caption{ Elastic constant, soft-mode flexocoupling coefficient and 
  correlation energy (self-correlation of the soft-mode branch) associated to a 
  transverse phonon propagating along the two directions considered in this work.
  Values are reported in eV. 
\label{tab7}}
\end{table}

With all the numerical data in our hands, we can now go back to the
continuum thermodynamic functional proposed in Section~\ref{sec:continuum},
and validate its accuracy against our reference first-principles model of
SrTiO$_3$. 
Given the lattice-dynamical nature of the formalism, comparing the 
phonon spectrum of the continuum Lagrangian with the corresponding 
first-principles dispersion curves appears as an excellent way
to assess the quality of the approximations that have been 
adopted so far.~\cite{Maranganti-07}

In Fig.~\ref{fig2} I plot the transverse and longitudinal acoustic 
phonon branches along [100] and [110] as predicted by the continuum
model, and the whole \emph{ab initio} phonon spectrum along the same 
directions in reciprocal space.
The first observation that one can make is that the continuum 
model does not seem to reproduce the first-principles results
very accurately: the agreement between the two breaks down 
only a short distance away from the zone center.
For slighty larger values of $q$, the continuum curves dip downwards 
and plunge below zero. (In fact, the restoring force associated to 
larger wavevectors becomes negative, leading to an imaginary
frequency and hence to an instability of the model.)
This behavior is common to both longitudinal and transverse
branches, although it is much more pronounced in the case
of the latter.
The fact that, by plugging the calculated values of the 
relevant coupling tensors into the strain-gradient functional
of Section~\ref{sec:continuum}, one obtains a pathological
behavior (i.e. a thermodynamically unstable model) is no
big surprise: The strain-gradient tensor ${\bf h}$ is 
systematically negative and enters the Hamiltonian
with the highest order in the wavevector $q$, $\mathcal{O}(q^4)$.
Such a ``sign'' issue is well known in the literature,
and seems to be a rather ubiquitous occurrence in the physics 
of many crystalline materials.~\cite{Maranganti-07}
%
%
%
What is, on the other hand, surprising is how serious 
the problem is in the present SrTiO$_3$ case:
Instabilities here occur unusually close to the zone center
(compared, e.g. to the cases that were reported in 
Ref.~\onlinecite{Maranganti-07}), which is a consequence of 
the strong coupling between the transverse soft-mode and 
acoustic branches.

Fixing this issue appears as a daunting task if one wishes to keep
working with the simple strain-gradient functional of Section~\ref{sec:continuum}.
As we have discussed at length in the previous Section, the 
fourth-order dispersion of the TA branches is dominated by the interaction 
with the ferroelectric soft mode. As a consequence of this interaction,
the SGE energy acquires a negative contribution that is inversely
proportional to the square of $\omega$, the transverse 
soft-mode frequency.
In phenomenological theories of ferroelectrics one typically assumes that
this frequency follows a critical temperature behavior as
\begin{equation}
\omega^{2} \propto (T-T_{\rm C}),
\end{equation}
where $T_{\rm C}$ is the Curie temperature; this means that the
continuum model of Sec.~\ref{sec:continuum} becomes unstable at a 
length scale $\xi$ that diverges as $(T-T_{\rm C})^{-1/2}$.
This appears difficult to fix in practical implementations;
plus, the adiabatic approximation that regards optical modes as ``fast''
variables becomes unjustified in a proximity of $T_{\rm C}$.
%
%
%

An obvious way to circumvent this issue consists in modifying the functional
of Sec.~\ref{sec:continuum} by promoting the soft mode to an independent degree 
of freedom, as it is commonly done in the ferroelectric literature.~\cite{Yudin-13} 
%
For example, specializing for simplicity~\footnote{The calculation of the full correlation 
  matrix ${\bf g}$ in
  a correct tensorial form presents some additional 
  subtleties regarding the treatment of the long-range electrostatics 
  and the gauge invariance; these would require a long digression in
  order to be adequately clarified.}
to a given transverse branch along a 
fixed propagation direction, one can write
\begin{eqnarray} 
\mathcal{L}(u,\dot u, \phi,\dot \phi) &=&  \frac{\rho_{\rm mass}}{2} ( \dot{\phi}^2 - \phi^2 \omega^2) + \frac{\rho_{\rm mass}}{2} |\dot{u}|^2 \nonumber \\
            && - \frac{1}{2} \mathcal{C} (u')^2 - \frac{1}{2 \Omega} g (\phi')^2 - \frac{1}{\Omega} f \phi' u' \nonumber \\ 
            && -\frac{1}{2} h^{\rm B} (u'')^2.
\label{newfunc}            
\end{eqnarray}
Here $\phi$ is the soft-mode amplitude, $g$ its correlation energy 
(see Section~\ref{sec:g6}), $f$ the corresponding flexocoupling 
coefficient, and $\mathcal{C}$ the elastic constant.
The contributions to the SGE energy that are not due to the soft 
mode have been grouped into the ``background'' SGE coefficient $h^{\rm B}$;
primed symbols refer to spatial derivatives along the propagation 
direction.
It is straightforward to show that the Lagrangian of Eq.~(\ref{newfunc})
reproduces, up to fourth order in $q$, the same dispersive behavior of the 
acoustic phonon branch as the simpler functional of Sec.~\ref{sec:continuum};
thus, the two formulations provide an equally accurate description of 
SGE effects.
%

To understand why the new functional is preferrable to that of Sec.~\ref{sec:continuum}
in the present SrTiO$_3$ case, in Fig.~\ref{fig2}(b) we present a lattice-dynamical 
analysis of the dispersion curves as calculated from Eq.~(\ref{newfunc}), 
either by including (dotted curves) or neglecting (dashed curves) the background
SGE term.
%
%
In the approximate ($h^{\rm B}=0$) version, which we shall discuss first, 
the instabilities have disappeared completely; this is a consequence
of suppressing the negative $\mathcal{O}(q^4)$ SGE contribution due to 
$h^{\rm B}$.
One can show that the resulting functional is thermodynamically stable at any
value of $\omega$ if
the (now highest) $\mathcal{O}(q^2)$ term is defined positive.
This requires the following condition~\cite{Axe-70,Yudin-13} to be satisfied along 
all directions in $q$-space,
\begin{equation}
\Omega  \mathcal{C} g> f^2.
\label{stability}
\end{equation}
In Table~\ref{tab7} I report
the values of the relevant parameters calculated in the present 
first-principles model of SrTiO$_3$ along [100] and [110];
the stability criterion, Eq.~(\ref{stability}), is clearly satisfied
in both cases.
Note that although there is no \emph{explicit} SGE term, the \emph{implicit} 
contribution of the soft mode to the SGE energy, which constitutes more 
than 90\% of the total, is correctly described via the flexocoupling 
term.
This observation explains the remarkable accuracy of the resulting acoustic 
dispersion curves [dashed curves in Fig.~\ref{fig2}(b)], especially along the 
[100] direction.

%
%
%

For several different reasons (e.g., to ensure the gauge invariance of the
theory, or to study physical phenomena where strain gradients are exceptionally
large~\cite{Arias-15}, or more simply in nonferroelectric materials), 
one may be interested in a more accurate (i.e. beyond the soft-mode
approximation) treatment of the SGE energy. If this is the case, it becomes 
necessary to reincorporate the background SGE effects that have been neglected 
in the last few paragraphs. 
%
The complete functional of Eq.~(\ref{newfunc}), with the
correct $h^{\rm B}$ coefficient included, yields the acoustic phonon 
branches that are shown as dotted curves 
in Fig.~\ref{fig2}(b).
%
%
While there are some improvements in the description of the dispersion
in a vicinity of the zone center, most clearly along the [110] direction,
the systematically negative sign of $h^{\rm B}$ brings us back to the
stability issues that we have already mentioned when commenting on panel (a).
%
[Note that the critical wavevector at which the instabilities occur is much larger
than in panel (a), since part of the SGE energy has been delegated to the 
flexocoupling term, and is now almost unsensitive to the soft-mode frequency, 
$\omega$.]
%

The fact that most contributions to the SGE energy are negative (and hence prone
to instabilities when incorporated in a continuum model) in most materials
-- the present results for SrTiO$_3$ are no exception -- 
was observed before,~\cite{Maranganti-07} and several
workarounds have been proposed over the years.~\cite{Askes-11}
%
A popular strategy consists in replacing the unstable strain 
gradients with stable
inertia or acceleration gradients.~\cite{Askes-11}
This way, the dispersion of an acoustic phonon 
branch along a given direction can be, in principle, adjusted 
to match the first-principles results even without introducing
an explicit SGE term.
When moving to the 3D case, however, it appears unlikely that 
one could replace the information contained in the SGE tensor 
entirely via this trick.
The SGE tensor, as we have shown in Sec.~\ref{sec:static},
describes both static and dynamic effects, and while inertia gradients 
may be used to reproduce the latter, they cannot obviously 
mimick the former. 
%

The concepts developed in this work naturally suggest two additional
strategies that could be used, as an alternative to (or in combination with) 
the inertia gradients, to construct thermodynamically stable SGE functionals.
%
%
The first, which would be ideally suited to a numerical 
implementation, consists in \emph{discretizing} the field equations,
e.g., via ``quasicontinuum'' methods.~\cite{Dupuy-05} Such
techniques have been successfully applied in the past to modeling the  
elastic properties of materials and nanostructures in a 
multiscale framework.~\cite{Colombo-10} Discretization naturally
introduces a low-pass filter in the (spatial) frequency spectrum
of the allowed solutions, and therefore looks particularly promising 
in the present context, where the problematic instabilities occur at 
exceptionally short length scales.
Moreover, such an approach is consistent with the physical
origin of, at least, part of the SGE energy (the frozen-ion contribution), 
which is precisely related to the discrete nature of the
atomic lattice (see Sec.~\ref{sec:disp}).

A second possibility involves incorporating an auxiliary
vector field in the continuum model, whose physical
parameters (zone-center frequency,
correlation and flexocoupling coefficient) 
%
are such that:
(i) the auxiliary mode is adiabatically
separated from both the soft mode and the acoustic branches;
(ii) its contribution to the SGE energy is equal or 
\emph{more negative} than any calculated $h^{\rm B}$ coefficient;
(iii) the stability condition Eq.~(\ref{stability}) is satisfied.
Given (i-iii), one is left then with a \emph{positive-defined} 
${\bf h}^{\rm B}$ tensor [the contribution from the auxiliary field,
in the form of Eq.~(\ref{hlm}), must be subtracted from ${\bf h}^{\rm B}$
in order to keep the overall SGE energy unaltered]. This implies that,
by introducing an additional degree of freedom in the model, and by carefully
engineering its (flexo)coupling to the deformation field, one can always
obtain a stable functional, and yet an exact (in relationship to the first-principles 
reference model) treatment of all the contributions to the SGE energy.

Exploring the details of such an approach (or of the discretization 
route that I have mentioned earlier) would bring me far from the
main scopes of this work, and I defer it to a future publication.
Still, the above discussion highlights the advantages of the strategy 
used in this work, i.e. of approaching continuum problems with a 
fundamental lattice-dynamical mindset. 
This way, one can not only extract realistic material-specific values of the
coefficients, but also provide firm microscopic foundations to the higher-level
theory, and possibly devise effective solutions to existing mathematical
puzzles.

\section{Conclusions}

\label{sec:conclusions}

I have derived a unified formulation of flexoelectricity and strain-gradient 
elasticity in crystalline insulators, and discussed its implications for 
(incipient) ferroelectric materials.
The ideas presented here are immeditely relevant to a vast range of physical 
phenomena involving spatial inhomogeneities in the strain or other order 
parameters.
Such studies have traditionally been the almost exclusive realm of phenomenological
approaches; this work clearly demonstrates that a fully \emph{ab initio} route,
via a hierachical multiscale framework, is a powerful (and very realistic) 
alternative.

In the present implementation, first-principles data have been used
as the ``exact'' reference on which the continuum model is built. One should 
keep in mind, in this context, that \emph{ab initio} approaches are not free
from limitations: on one hand, there are the well-known accuracy concerns related
to the approximate treatment of the exchange and correlation energy; on the other
hand, direct electronic-structure methods are only practical at zero temperature,
which at first sight thwarts their applicability to the accurate study of ferroelectric
materials.
Neither of the above two issues is, in fact, a drawback of the method described here.
The present multiscale strategy is completely general, and readily applicable to
an arbitrary microscopic model. (This can be either an \emph{ab initio} 
or a classical atomistic description.) Furthermore, the long-wave approximation,
combined with the quasicontinuum approach that I have mentioned in Sec.~\ref{sec:discuss},
can be regarded as a powerful, systematic tool to construct \emph{effective Hamiltonian}~\cite{heff,zhong/vanderbilt/rabe:1995,ponomareva}
models. 
The latter have been successfully used during the past two decades as a means to
exploring finite-temperature and other effects in complex ferroelectric systems.
In this respect, this work may open interesting new avenues towards overcoming
the stringent time- and length-scale limitations of direct \emph{ab initio} 
approaches; exploring these opportunities will be an interesting topic for
future studies.

Based on the above considerations, I expect this work to promote a closer synergy 
between condensed-matter theorists that are active in the field of continuum 
modeling with those in the first-principles community, with many exciting
opportunities for future collaboration.
Apart from the obvious application to flexoelectricity, the methodologies
developed here are directly relevant to ferroelectrics at large, and
more generally to any physical system where electrical
and mechanical degrees of freedom couple in nontrivial ways.

%

\section*{Acknowledgments}


This work was supported 
by MINECO-Spain through Grants No. FIS2013-48668-C2-2-P
and No. SEV-2015-0496, and Generalitat de Catalunya (2014 SGR301).
Calculations were performed at Supercomputing Center of Galicia (CESGA).

\appendix*

\bibliography{merged}

\end{document}